\documentclass [prb,reprint,superscriptaddress,floatfix] {revtex4-1}
\usepackage{amsmath}
\usepackage{graphicx}
\usepackage{lmodern}
\usepackage{amsmath}

\usepackage{color}
\usepackage{amssymb}
\newcommand{\beq} {\begin{equation}}
\newcommand{\eeq} {\end{equation}}
\newcommand{\bea} {\begin{eqnarray}}
\newcommand{\eea} {\end{eqnarray}}
\newcommand{\be} {\begin{equation}}
\newcommand{\ee} {\end{equation}}
\renewcommand{\(}{\left(}
\renewcommand{\)}{\right)}
\renewcommand{\[}{\left[}
\renewcommand{\]}{\right]}
\newcommand{\TRS}{time-reversal symmetry }
\newcommand{\TRSB}{time-reversal symmetry breaking }
\newcommand{\oncite}[1]{Ref.\ [\onlinecite{#1}]}
\DeclareMathOperator{\sgn}{sgn}
\DeclareMathOperator{\Tr}{Tr}

\DeclareMathOperator{\Imm}{Im}

\begin{document}

\title {Polar Kerr effect from chiral-nematic charge order}
\author{Yuxuan Wang}
\affiliation{Department of Physics, University of Wisconsin, Madison, WI 53706, USA}
\author{Andrey Chubukov}
\affiliation{Department of Physics, University of Wisconsin, Madison, WI 53706, USA}
\affiliation{William I. Fine Theoretical Physics Institute,
and School of Physics and Astronomy,
University of Minnesota, Minneapolis, MN 55455, USA}
\author{Rahul Nandkishore}
\affiliation{Princeton Center for Theoretical Science, Princeton University, Princeton, NJ 08544, USA}
\date{\today}

\begin{abstract}
We analyze the polar Kerr effect in an itinerant electron system on a square lattice in the presence of a composite charge order proposed for the pseudogap state in underdoped cuprates. This composite charge order preserves discrete translational symmetries, and is ``chiral-nematic" in the sense that it breaks time-reversal symmetry, mirror symmetries in $x$ and $y$ directions, and $C_4$ lattice rotation symmetry.
 The Kerr angle $\theta_K$ in $C_4$-symmetric system is proportional to the antisymmetric component of the anomalous Hall conductivity $\sigma_{xy}-\sigma_{yx}$. We
   show that this result holds when $C_4$ symmetry is broken. We show that in order for $\sigma_{xy}$ and $\sigma_{yx}$ to be non-zero the mirror symmetries in $x$ and $y$ directions have to be broken, and that for $\sigma_{xy}-\sigma_{yx}$ to be non-zero time-reversal symmetry has to be broken. The chiral-nematic charge order satisfies all these conditions, such that a non-zero signal in a polar Kerr effect experiment is symmetry allowed.
We further show that to get a non-zero $\theta_K$ in a one-band spin-fluctuation scenario, in the absence of disorder, one has to extend the spin-mediated interaction to momenta away from $(\pi,\pi)$ and has to include particle-hole asymmetry. Alternatively, in the presence of disorder one can get a non-zero $\theta_K$ from impurity scattering: either due to skew scattering (with non-Gaussian disorder) or due to particle-hole asymmetry in case of Gaussian disorder.  The impurity analysis in our case is similar to that in earlier works on Kerr effect in $p_x+ip_y$ superconductor, however in our case the magnitude of $\theta_K$ is enhanced by the flattening of the Fermi surface in the
``hot" regions which mostly contribute to charge order.
\end{abstract}
\maketitle

\section{Introduction}
The analysis of the pseudogap phase in underdoped cuprates remains a hot topic in research on correlated electron systems. The two
 generic competing scenarios associate the pseudogap with (i) precursor behavior to either Mott physics, antiferromagnetism, or both~\cite{anderson,lee,rice,millis,tremblay_1,ph_ph}
 and (ii) competition between superconductivity and competing order, e.g., loop current order~\cite{varma} or more conventional
 charge-density-wave (CDW) order with form factors of different symmetries~\cite{castellani,ddw,kotliar,ms,efetov,charge}.
Gross features of the pseudogap, like  re-distribution of the spectral weight from low to higher energies, can be understood within both scenarios, leaving a possibility that both contribute to the physics of the pseudogap and the interplay between the two depends on the doping. More subtle features, discovered over the past few years in intensive experimental studies of underdoped cuprates, appear to support the competing order scenario in the sense that the data show that the the pseudogap is likely a phase (or even a set of phases) with broken symmetries.
       The evidence for broken symmetries comes from different sources.
       First, two sets of experiments indicate that time-reversal symmetry may be broken.
       One is the observation of the
       polar Kerr effect
       in YBa${_2}$Cu${_3}$O$_{6+x}$ (YBCO)~\cite{YBCO_kerr} and La$_{1.875}$Ba$_{0.125}$CuO$_4$ (LBCO)~\cite{LBCO_kerr}
        below some
       critical temperature $T_K(x)$, which increases as $x$ decreases.
         Another is the observation of intra-unit-cell magnetic order in polarized neutron scattering measurements
  \cite{bourges,greven}.
    The onset temperature of
    the intra-unit-cell order  is not the same as $T_K$, but roughly follows the same doping dependence.
Second, recent optical experiments in the terahertz regime have found~\cite{armitage} a non-zero linear birefringence, which was interpreted as the result of the breaking of $C_4$ lattice rotational symmetry. The tilt of the pattern of dichroism was additionally related to the breaking of mirror symmetries.
X-ray and neutron scattering data on LBCO (Refs.\ [\onlinecite{labacuo}, \onlinecite{stripes}]), neutron scattering in YBCO (\oncite{hinkov}), STM data on Bi$_2$Sr$_2$CaCu$_2$O$_{8+\delta}$
(Refs.\ [\onlinecite{davis}, \onlinecite{davis_1}]), and measurements of
 longitudinal thermo-electric coefficient in YBCO (\oncite{taillefer_last})
 also indicate that lattice rotational symmetry is broken from $C_4$ down to $C_2$ below a certain temperature
 comparable to $T_K$.
Third,
STM, resonance X-ray, NMR, and other measurements found evidence for charge order
in the underdoped regime of several families of the cuprates~\cite{davis_1,ybco, ybco_1, X-ray, X-ray_1,wu,mark,ultra,suchitra,zxshen,zxshen_0}.
 Measurements of Hall and Seebeck coefficients~\cite{taillefer} were interpreted as feedback from the CDW order on the fermions.
   The charge order sets in at incommensurate momenta $Q_x = (Q,0)$ and $Q_y=(0, Q)$ and has a predominantly $d$-wave form factor~\cite{davis_1}.
    The observed order is static, yet a short-range one.
     Whether the latter is an intrinsic phenomenon, or due to pinning of would-be long-range order by
     impurities  remains to be seen~\cite{steve_last,julien_last}.

An incommensurate charge order with momentum $Q_x/Q_y$ breaks translational (phase) $U(1)$
symmetry, but  can also break two discrete $Z_2$ symmetries.
One is $C_4$ lattice rotational symmetry. It is broken down to $C_2$ if the order develops with momentum $Q_x$ or $Q_y$, but not both.
Another is time-reversal symmetry. Its potential breaking is associated with the fact that for
$Q$
      extracted from the data\cite{X-ray}, the CDW order $\Delta^Q_k = \langle c^\dagger_{k+Q/2} c_{k-Q/2}\rangle$, involves low-energy fermions
      at $k+Q/2$ and $k-Q/2$ with the center-of-mass momentum  $k$  not located at any special symmetry point in the Brillouin zone 
 ($k=\pm k_0$ or $\pm k_0+(\pi,\pi)$ in Fig.\ \ref{fig:1}),
  Accordingly, while $|\Delta^Q_k| =|\Delta^Q_{-k}|$, the phases of the two $U(1)$ order parameters $\Delta^Q_k = |\Delta^Q_k| e^{i\phi_k}$ and $\Delta^Q_{-k} =|\Delta^Q_{-k}| e^{i\phi_{-k}}$
 are in general not identical. The CDW orders $\Delta^Q_k$ and $\Delta^Q_{-k}$ are related by time reversal operation~\cite{charge,laplaca,agterberg},
 hence if $\phi_k$ and $\phi_{-k}$ are different, time reversal transforms a  state with a given $\delta \phi = \phi_k -\phi_{-k}$ into a different state with
 $-\delta \phi$.  Note that this breaking of time-reversal symmetry is specific to CDW order with $Q_x$ or $Q_y$.
 (i.e.
 ${\bf Q}$ along one of the two symmetry axes).
 For CDW order with momenta $(Q, \pm Q)$ along a
 zone diagonal, $k$ necessarily equals to $(\pi,0) \equiv (-\pi,0)$, hence $\Delta^Q_k$ and $\Delta^Q_{-k}$ are identical~\cite{charge}.

 At the mean-field level the continuous $U(1)$ symmetry and the two discrete $Z_2$ symmetries get broken at the same temperature $T$.
 Beyond mean-field, the two $Z_2$ symmetries get broken prior to the breaking of the $U(1)$ symmetry.  In the intermediate regime
 the system develops a nematic order by choosing $Q=Q_x$ or $Q_y$ and breaks time-reversal symmetry by choosing the relative phase of, say, $\Delta_k^{Q_x}$ and $\Delta_{-k}^{Q_x}$
 to be $\delta\phi$ or $-\delta\phi$
 , however the common phase of $\Delta_k$ and $\Delta_{-k}$ remains unordered.
 Such a state
  does not break $U(1)$ phase symmetry and does not have
  a two-fermion
  CDW
  condensate i.e., $\langle c^\dagger_{k+Q/2} c_{k-Q/2}\rangle =0$, but composite order parameters consisting of four fermionic operators, develop
  non-zero expectation values. 
  The order parameter for the nematic order is Ising variable $\Gamma = \langle|\Delta_{\pm k}^{Q_x}|^2 -|\Delta_{\pm k}^{Q_y}|^2\rangle$,
  and the one for the time-reversal symmetry breaking is Ising variable $\Upsilon = i \langle \Delta_{k}^Q(\Delta_{-k}^Q)^* - \Delta_{-k}^Q(\Delta_k^Q)^*\rangle\equiv-2|\Delta_k^Q|^2\sin\delta\phi$, where $Q =Q_x$ or $Q_y$. The appearance of a non-zero $\Upsilon$ also implies that mirror symmetries along $x$ and $y$ directions are broken because for, e.g. $Q=Q_x$,
 mirror reflection along $y$-direction changes $\Delta_k\leftrightarrow\Delta_{-k}$ and mirror reflection along $x$-direction changes $\Delta_{\pm k}\to\Delta_{\pm k}^*$.
  Under both reflections, $\Upsilon$ changes to $- \Upsilon$, just as it does under time reversal.
  In real space, this composite order implies that  incommensurate charge and current modulations fluctuate at any given ${\bf r}$
  and average to zero, yet the anti-correlation between the two modulations remains the same as in the absence of fluctuations.

  It was argued in Refs.\ [\onlinecite{charge},\onlinecite{tsvelik}] that such a partly ordered state already leads to pseudogap behavior and is a candidate for the pseudogap phase in the
   temperature/doping range where CDW order does not yet develop. The goal of the present communication is to analyze whether such $Z_2 \times Z_2$ order  gives rise to a non-zero Kerr effect.  We show that it does. Once the Kerr signal develops in the phase with a composite order, it remains non-zero  also in the parameter range when the system develops a true CDW order
    simply because $Z_2 \times Z_2$
     remains
      a part of the order parameter manifold.

     Polar Kerr effect measures the rotation angle $\theta_K$ of the polarization plane of light upon normal reflection from the
  surface. It has been used as a probe for \TRSB in $p$-wave superconductor Sr$_2$RuO$_4$ (\oncite{sro}) and heavy fermion superconductor UPt$_3$ (\oncite{upt}).
Despite a number of theoretical attempts, the nature of Kerr effect in cuprates remains controversial. It is generally expected that the external applied magnetic field should be able to train the sign of the Kerr angle. Experiments on, e.g., Sr$_2$RuO$_4$ do show the switch of the sign of $\theta_K$ under the
  change of the direction of the applied field. However, in the cuprates no such training has been observed in fields up to 14 Tesla.
   Furthermore, there is no sign change of Kerr signal upon flipping the sample.
   The absence of the training effect by a field has inspired efforts to explain the Kerr effect in cuprates by invoking a symmetry breaking different from  time-reversal, e.g. a gyrotropic order~\cite{raghu, orenstein, varma2,helical}. However, it has been recently  pointed out~\cite{raghu_corr,helical_corr} that in a system that satisfies the Onsager relations and is in the linear response regime (when the current is proportional to the electric field) \TRSB is a necessary precondition
    for the observation of a non-zero Kerr response.  It is possible
      in principle
      that Kerr effect in the cuprates may be caused by extrinsic time reversal symmetry breaking such as non-equilibrium effects and non-linear response~\cite{raghu_corr}. However, why cuprates would be especially susceptible to these effects remains to be elucidated.

In this paper we investigate whether a Kerr
 signal
 consistent with experiments can emerge from a chiral-nematic charge order in the linear regime. We show that a non-zero Kerr rotation is symmetry allowed. We also show that in model calculations to get a non-zero $\theta_k$ one has to either invoke
   particle-hole asymmetry or, if particle-hole symmetry is approximately preserved, include non-Gaussian-type disorder (a skew scattering).
     In this respect, the situation is similar to  earlier studies of the Kerr effect in  a $p_x+ip_y$ superconductor (Refs. [\onlinecite{goryo,yakovenko,mineev,kallin}]).
 However, in square lattice itinerant fermion models the Kerr effect is enhanced because of the nearly flat Fermi surface in the antinodal regions. We
   argue
   that the Kerr signal from a chiral-nematic charge order should not be susceptible to training with an external magnetic field. This is because in real space the order parameter $\Upsilon$  induces a magnetic field that oscillates in space and averages to zero when integrated over the system area.
  Such induced magnetic field does not couple linearly  to a uniform external training field.
   Our analysis does not explain the absence of the sign change of the Kerr signal when the sample is flipped. In fact, for any two-dimensional system, the Kerr signal should {\it by definition} reverse sign upon flipping the sample.
   It is possible that the resolution lies in the interlayer coupling between the CuO$_2$ planes~\cite{akash} and that the discrepancy may be resolved by studying the three-dimensional structure of the \TRSB  order parameter~\cite{raghu, orenstein, varma2,helical,yakovenko_future}.
 This 3D analysis is  beyond the scope of the present paper. We hope that our study in a single two-dimensional layer will provide a basis for a future analysis
  of the Kerr effect in three-dimensional systems~\cite{yakovenko_future}.

The outline of our paper is the following. In Sec.\ \ref{model} we briefly outline the microscopic model for the order parameters.
In Sec.\ \ref{cal} we derive $\theta_K$ for our system with the composite order $\Upsilon$.
We first show (Sec.\ \ref{sym})
 that the symmetry requirements are satisfied by this composite order.
We
  then
  detail the calculation of $\theta_K$ for the case of impurity scattering
  (Sec.\ \ref{imp}), and for the case of spin fluctuations
  (Sec.\ \ref{spf})
  and
  discuss the effect of the Fermi surface geometry (Sec.\ \ref{fsg}).
  In Sec.\ \ref{disc} we discuss
   how Kerr signal is affected by
    applying a uniform magnetic field and flipping the sample.
   We present our conclusions in Sec.\ \ref{conc}.
    In our analysis we do make use of the fact that the Kerr angle $\theta_K$ is related to the antisymmetric part of the Hall conductivity $\sigma_{xy}-\sigma_{yx}$, which is non-zero only then time-reversal symmetry is broken. This result seems intuitive, yet
 to obtain it in the absence of $C_4$ symmetry we need to invoke the details of the design of the experiments which measure $\theta_K$~\cite{LBCO_kerr,kapitulnik,xia}. This discussion is presented in the Appendix
 along with some technical details of the calculations.

\section{Structure of the charge order}
\label{model}
\begin{figure}
\includegraphics[width=.8\columnwidth]{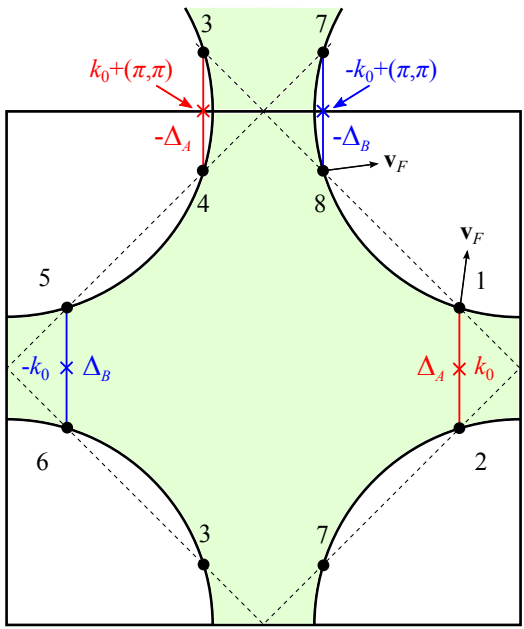}
\caption{The Fermi surface in the first Brillouin zone, showing eight hot spots (labeled 1-8) and charge-density-wave (CDW) pairings between them (denoted as ``bonds" connecting hot spots).
The momentum ${\bf Q}$ is chosen along $y$ direction: ${\bf Q} = Q_y = (0,Q)$.
 The CDW order parameter differs by a sign when shifted by momentum $(\pi,\pi)$. $\Delta_A$ and $\Delta_B$ are related by mirror reflection and time reversal. The Fermi velocities at hot spots 1,2,5,6 are almost along $y$ direction, and at hot spots 3,4,7,8 they are almost along $x$ direction.
The momentum ${\bf k}_0$
is along $x$ direction for ${\bf Q} = Q_y$, i.e., ${\bf k}_0 = (k_0,0)$. The value $k_0$ is $\pi-Q/2$.
   }
\label{fig:1}
\end{figure}

We follow earlier works~\cite{ms,efetov,charge,tsvelik,efetov_2,efetov_3,subir_3,subir_4,bill} and assume that CDW order and chiral-nematic charge order which precedes it
emerge due to magnetically mediated interaction between fermions in ``hot regions"  around eight hot spots, which are defined~\cite{acs} as intersections between the Fermi surface and magnetic Brillouin zone boundary. We label these hot spots by number 1-8 in Fig.\ \ref{fig:1}. For
hole-doped cuprates, the Fermi surface is ``flattened" in the near anti-nodal regions, i.e., at hot spots the fermi velocities are either nearly vertical ($v_F\sim v_y\gg v_x$ for hot spots 1,2,5,6) or nearly horizontal ($v_F\sim v_x\gg v_y$ for hot spots 3,4,7,8).

Two of us recently analyzed~\cite{charge,tsvelik}
  the structure of CDW order with $Q_x/Q_y$ within a spin-fluctuation scenario,
   using spin-fermion model as input and
  using the Hubbard-Stratonovich formalism to derive the
  Ginzburg-Landau free energy for four  $U(1)$ order parameters $\Delta_{A}^{x}$, $\Delta_{B}^{x}$, $\Delta_{A}^y$ and $\Delta_{B}^y$,
  where $\Delta_{A,B}^{x,y}\equiv\Delta_{\pm k_0}^{Q_{x,y}}$ ($\Delta_A^y$ and $\Delta_B^y$ are shown in
  Fig.\ \ref{fig:1}). 
  We assume that each order parameter is a constant within a small center-of-mass momentum range around the corresponding hot spots.
  Each $U(1)$ order parameter is a complex field with amplitude and phase.
    The order parameters at the center of mass momenta $\pm ({\bf k} + (\pi,\pi))$ have about the same amplitude as the ones $\pm {\bf k}$ but the phase is shifted
     by $\pi$, i.e.,  $\Delta_{k+(\pi,\pi)}^{i} \approx - \Delta_{k}^{i}$.

 The analysis of Ginzburg-Landau functional has demonstrated that the most likely CDW state is the one in which (i) both $\Delta_A$ and $\Delta_B$ develop with either $Q_x$ or $Q_y$ but not both, breaking a $Z_2$ lattice rotation symmetry (the stripe order), and
  (ii) for a given choice of ${\bf Q}$, say $Q_y$, the minimization of free energy prefers a phase difference $\pm\delta\phi$ between $\Delta_A$ and $\Delta_B$, where $\delta\phi$ is a fixed number. By specifying
the sign for $\pm\delta\phi$ the system breaks an additional $Z_2$ symmetry, which in our case is time-reversal.
In the CDW-ordered phase.
the system also spontaneously breaks the $U(1)$ translational symmetry by choosing some value of the common phase of $\Delta_A^y$ and $\Delta_B^y$.
 In real space a non-zero $\langle\Delta_A\rangle$ and $\langle\Delta_B\rangle$
  create an incommensurate charge
  modulation measured by $\langle\Delta_A\rangle+\langle\Delta_B\rangle$ (predominantly bond order for $\Delta_{k+(\pi,\pi)}^{i} \approx - \Delta_{k}^{i}$)
    and incommensurate current modulation measured by $\langle\Delta_A\rangle- \langle\Delta_B\rangle$. Since $|\langle\Delta_A\rangle|=|\langle\Delta_B\rangle|$, the two modulations are shifted in phase by $\pm \pi/2$
(i.e., in real space, if one is $\cos{{\bf Qr}}$, another is $\sin{{\bf Qr}}$).

We use these results as input and consider
  a CDW order $\Delta_{k}^Q\equiv\langle c^\dagger_{k+Q/2} c_{k-Q/2}\rangle$ with $Q=Q_y$, i.e., assume that $C_4$ symmetry is already broken down to $C_2$.
Without loss of generality, we
assume this CDW order has purely a
 $d$-form factor, namely $\Delta_k^Q=-\Delta_{k+(\pi,\pi)}^Q$. The CDW order parameter is peaked strongly at hot spots and we denote the center-of-mass momenta of the relevant hot spot pairs as $\pm k_0$ and $\pm k_0+(\pi,\pi)$ respectively (see Fig.\ \ref{fig:1}).
   In Table \ref{op}, we list the relevant bilinear fermionic operators and their expectation values.
   \begin{table}
 \caption{The expectation values of the CDW operators between hot spots 1-8.}
 \begin{ruledtabular}
 \begin{tabular}{cccc}
 $\langle c_1^\dagger c_2\rangle$ & $\langle c_3^\dagger c_4\rangle$ & $\langle c_5^\dagger c_6\rangle$ & $\langle c_7^\dagger c_8\rangle$  \\
 $ \Delta_A$ & $-\Delta_A$ & $\Delta_B$ & $-\Delta_B$ \\ \hline
  $\langle c_2^\dagger c_1\rangle$ & $\langle c_4^\dagger c_3\rangle$ & $\langle c_6^\dagger c_5\rangle$ & $\langle c_8^\dagger c_7\rangle$  \\
 $ \Delta_A^*$ & $-\Delta_A^*$ & $\Delta_B^*$ & $-\Delta_B^*$ \\ \end{tabular}
 \end{ruledtabular}
 \label{op}
 \end{table}
The CDW order parameters introduce anomalous vertices $\Delta_A(c_2^\dagger c_1-c_4^\dagger c_3)+h.c.$ and $\Delta_B(c_6^\dagger c_5-c_8^\dagger c_7)+h.c.$ to the effective action.
Throughout this work, we assume that we are above the superconducting temperature $T_{\rm sc}$, such that superconducting order does not come into play.

The order parameter which describes the breaking of $Z_2$ time-reversal symmetry (mirror symmetries in $x$ and $y$ directions) is
$\Upsilon = -i \langle\Delta_A^*\Delta_B-\Delta_B^*\Delta_A\rangle$.
 As we said in the Introduction, $Z_2$ symmetry associated with  lattice rotational $C_4$ symmetry  $Z_2$  associated
  with time-reversal/mirror symmetries get broken at a higher temperature than the one when $U(1)$ phase symmetry gets broken and  $\langle\Delta_A\rangle$ and $\langle\Delta_B\rangle$ individually become non-zero. Below we consider the Kerr effect in the $T$ range where the two $Z_2$ symmetries are broken but $U(1)$ symmetry is still preserved.
  We will see that the breaking of $Z_2$ lattice rotational symmetry does not influence the Kerr effect, while breaking of $Z_2$ time-reversal/mirror symmetry
    plays a crucial role.

\section{The calculation of Kerr angle}
\label{cal}
In the $C_4$-symmetric system the Kerr angle is related to the antisymmetric part of the Hall conductivity by~\cite{kerr_angle}
\begin{align}
\theta_K=\frac{\lambda}{c}\Imm\[\frac{\sigma_{xy}-\sigma_{yx}}{n(n^2-1)}\],
\label{9}
\end{align}
where $\lambda$ is the wavelength, $c$ the speed of light, and $n$ the refractive index.
When $C_4$ symmetry is broken, the relation between optical conductivity and Kerr angle is in general more complex (see Appendix \ref{em}).
 We found, however, that for the experimental setup used by Kapitulnik group \cite{kapitulnik,xia}, Eq.\ (\ref{9}) still holds, i.e.,
 $\theta_K$ is still directly proportional to $\sigma_{xy}-\sigma_{yx}$.

The Hall conductivity is calculated via the Kubo formula
\begin{align}
\sigma_{xy}(\omega)=\frac{1}{i\omega d}Q_{xy}^R(q=0,\omega),
\label{kubo}
\end{align}
where $\omega$ is the frequency of the incident light. The factor $d$ is the distance between neighboring CuO$_2$ layers
 The  retarded current-current correlator is defined as
\begin{align}
Q_{xy}^R(q=0,\omega)\equiv \int_0^{\infty} dt e^{i\omega t}\langle [\hat j_x(q=0, t),\hat j_y(q=0, 0)]\rangle,
\label{ret}
\end{align}
where $\hat j_x\equiv \psi^{\dagger}\hat v_x \psi$ and $\hat j_y\equiv \psi^{\dagger}\hat v_y \psi$ are the current operators, and $\hat v_x$ and $\hat v_y$ are velocity operators in the momentum space.

\subsection{Symmetry requirements}
\label{sym}
Before we calculate the Kerr angle, we first look at the symmetry properties of the Hall conductivities.

 Under mirror reflection along $x$ ($y$) direction, the current $j_{y(x)}$ transforms into $-j_{y(x)}$. From Eqs.\ (\ref{kubo}) and (\ref{ret}), we find that $\sigma_{xy}(\omega)$ and $\sigma_{yx}(\omega)$ transforms to $-\sigma_{xy}(\omega)$ and $-\sigma_{yx}(\omega)$. Time reversal is a bit more tricky. Generally under time reversal we have,
  \begin{align}
\langle\alpha|\hat O|\beta\rangle\to\langle\beta|(\hat T^{-1}\hat O\hat T)^{\dagger}|\alpha\rangle,
\label{tr}
\end{align}
 where $\hat T$ is an anti-unitary operator: $\hat T=\hat U\hat K$, where $\hat K$ is complex conjugation operator and $\hat U$ reverses momentum and flips spin. Using this relation, we find that under time reversal,
 \begin{align}
& \sigma_{xy}(\omega)\nonumber\\
&\to\frac{1}{i\omega d}\int_{0}^\infty dte^{i\omega t}\langle( \hat T^{-1}[\hat j_x(q=0, t),\hat j_y(q=0, 0)]\hat T)^\dagger\rangle,\nonumber\\
&=\frac{1}{i\omega d}\int_{0}^\infty dte^{i\omega t}\langle [\hat j_y(q=0, 0),\hat j_x(q=0, -t)]\rangle\nonumber\\
&=\sigma_{yx}(\omega).
 \end{align}
  We list these transformations in Table \ref{tab1}.

  \begin{table}
 \caption{The symmetry properties of the Hall conductivities under mirror reflections ${\cal M}_x$ and ${\cal M}_y$ and time reversal $\cal T$.}
 \begin{ruledtabular}
 \begin{tabular}{cccc}
 &${\cal M}_x$ & ${\cal M}_y$ & ${\cal T}$  \\ \hline
 $\sigma_{xy}(\omega)$ & $-\sigma_{xy}(\omega)$ & $-\sigma_{xy}(\omega)$ & $\sigma_{yx}(\omega)$ \\
 $\sigma_{yx}(\omega)$ & $-\sigma_{yx}(\omega)$ & $-\sigma_{yx}(\omega)$ & $\sigma_{xy}(\omega)$ \\
 $\sigma_{xy}(\omega)-\sigma_{yx}(\omega)\propto\theta_K$ &  \multicolumn{3}{c}{$\sigma_{yx}(\omega)-\sigma_{xy}(\omega)$ } \\
 \end{tabular}
 \end{ruledtabular}
 \label{tab1}
 \end{table}

    \begin{table}
 \caption{The symmetry properties of the order parameters $\Delta_A$, $\Delta_B$ and $\Upsilon$ under mirror reflections ${\cal M}_x$ and ${\cal M}_y$ and time reversal $\cal T$.}
 \begin{ruledtabular}
 \begin{tabular}{cccc}
 &${\cal M}_x$ & ${\cal M}_y$ & ${\cal T}$  \\ \hline
 $\Delta_A$ & $\Delta_A^*$ & $\Delta_B$ & $\Delta_B$ \\
  $\Delta_B$ & $\Delta_B^*$ & $\Delta_A$ & $\Delta_A$ \\
 $\Upsilon=-i(\Delta_A\Delta_B^*-\Delta_B\Delta_A^*)$ & {$-\Upsilon$ } &{$-\Upsilon$ } & {$-\Upsilon$ }\\
 \end{tabular}
 \end{ruledtabular}
 \label{tab2}
 \end{table}

 From the last line of Table \ref{tab1}, We
 find that  under all these symmetry operations (mirror reflections and time-reversal)  $\sigma_{xy}(\omega)-\sigma_{yx}(\omega)$ transforms to  $\sigma_{yx}(\omega)-\sigma_{xy}(\omega)$. This has been discussed before~\cite{vanderbit,kleiner}. Hence,
 if {\it any} of the three symmetries is intact, then
  $\sigma_{xy}(\omega)-\sigma_{yx}(\omega)=
  0$. 
 Furthermore, since $\sigma_{xy}-\sigma_{yx}$ is invariant under arbitrary $SO(2)$ spatial rotation, the choice of $x$ and $y$ above can be arbitrary.
  Therefore a non-zero Kerr effect requires {\it all} mirror symmetries {\it and} \TRS to be broken.
  We emphasize that breaking of \TRS alone is not sufficient for obtaining a non-zero Kerr angle -- all the mirror symmetries must be broken also.

For the system we consider, lattice rotational symmetry $C_4$ is broken and all mirror symmetries are trivially broken, except for those along two lattice directions. We will only discuss mirror symmetries along these directions for our system and denote these two directions as $x$ and $y$ hereafter. We do caution, however, for a generic square lattice system, only breaking mirror symmetries along lattice directions is not enough, as there may exist mirror planes other than lattice directions such as $x\pm y$~\cite{varma2,yakovenko_future}. Indeed, we checked that if CDW were to develop along both $x$ and $y$ directions such that mirror symmetries along $x\pm y$ directions were intact, the Kerr effect contributed by CDW along $x$ and $y$ directions would cancel out, as dictated by symmetry.

Under mirror reflection along $y$-direction, hot spots $(1,2)\leftrightarrow(5,6)$, and $(3,4)\leftrightarrow(7,8)$ (see Fig.\ \ref{fig:1}). From Table \ref{op}, we find that $\Delta_A\leftrightarrow\Delta_B$. Under mirror reflection along $x$-direciton, hot spots $(1,2)\to (2,1)$ and $(3,4)\to (4,3)$. Again from Table \ref{op}, we find that $\Delta_{A,B}\to\Delta_{A,B}^*$. Under time reversal, all momenta are reversed, e.g. hot spots (1,2) transform into (6,5). From Eq.\ (\ref{tr}) we find that,
\begin{align}
\langle0| c_1^\dagger c_20|\rangle&\to\langle0|(\hat T^{-1}c_1^\dagger c_2\hat T)^\dagger|0\rangle\nonumber\\
&=\langle0|(c_6^\dagger c_5)^\dagger|0\rangle\nonumber\\
&=\langle0|c_5^\dagger c_6|0\rangle.
\end{align}
Proceeding similarly for other hot spot pairs in Table \ref{op}, we see that $\Delta_A\leftrightarrow\Delta_B$ under time reversal. We
 see therefore
 that under all symmetry operations above, our chiral-nematic charge order parameter $\Upsilon=-i(\Delta_A\Delta_B^*-\Delta_B\Delta_A^*)$ transforms into $-\Upsilon$, thus breaking all three symmetries. Therefore, a non-zero Kerr angle is symmetry allowed. For clarity, we list the symmetry properties of order parameters in Table \ref{tab2}.\\

Symmetry analysis shows that $\theta_K$ must scale with $\Upsilon$. To  obtain the proportionality coefficient we need to turn to microscopic
analysis. This is what we do next. We first consider impurity scattering and then scattering by spin fluctuations.
In both cases, we initially assume that a true CDW order exists, introduce
 the primary CDW order parameters $\Delta_A$ and $\Delta_B$ and express $\theta_K$ in terms of $\Delta_A^*\Delta_B-\Delta_B^*\Delta_A = i \Upsilon$.
  We then argue that  the relation between $\theta_K$ and $\Upsilon$ holds even when individually
   $\langle\Delta_A\rangle=\langle\Delta_B\rangle=0$.

 \subsection{Kerr effect due to disorder}
 \label{imp}
 \begin{figure}[htbp]
\includegraphics[width=\columnwidth]{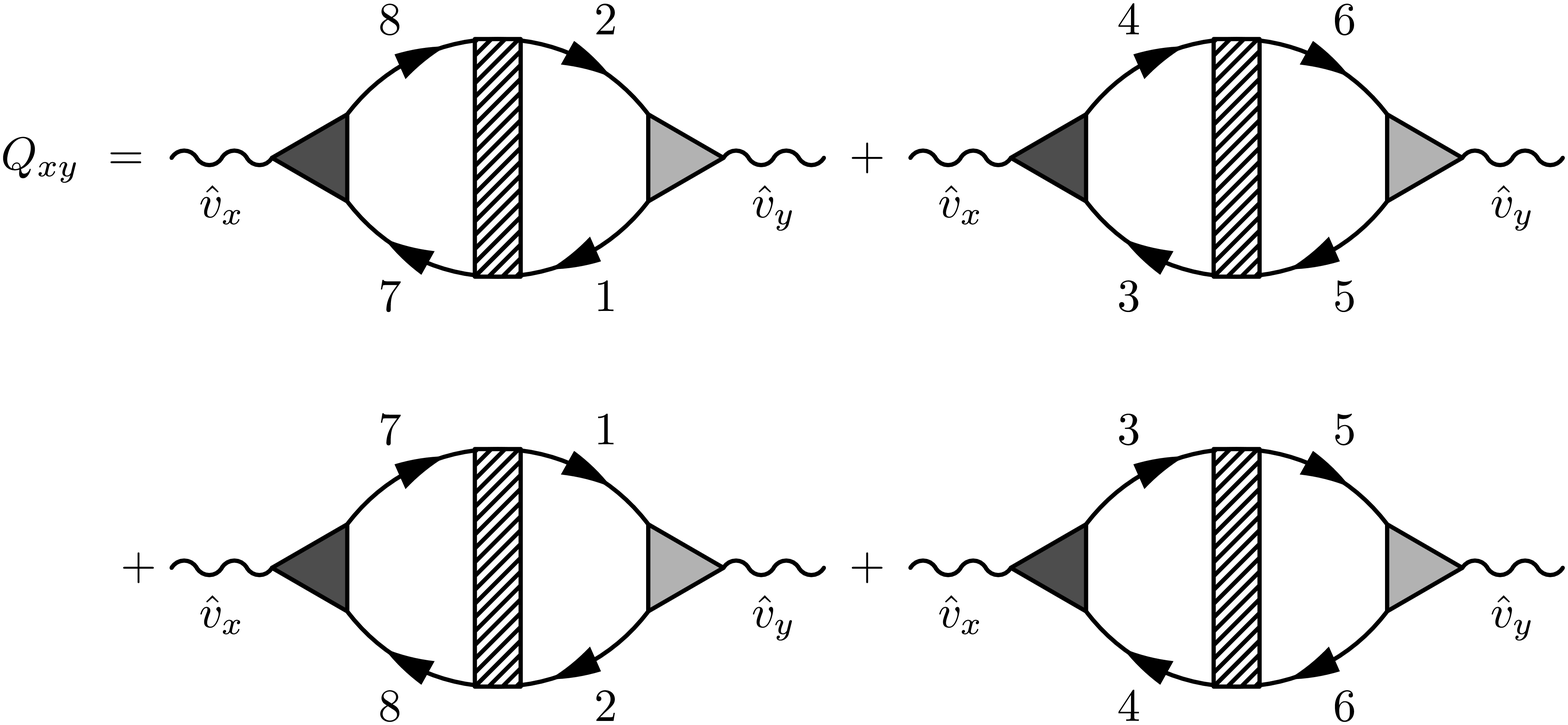}\vspace{0.3cm}
\includegraphics[width=\columnwidth]{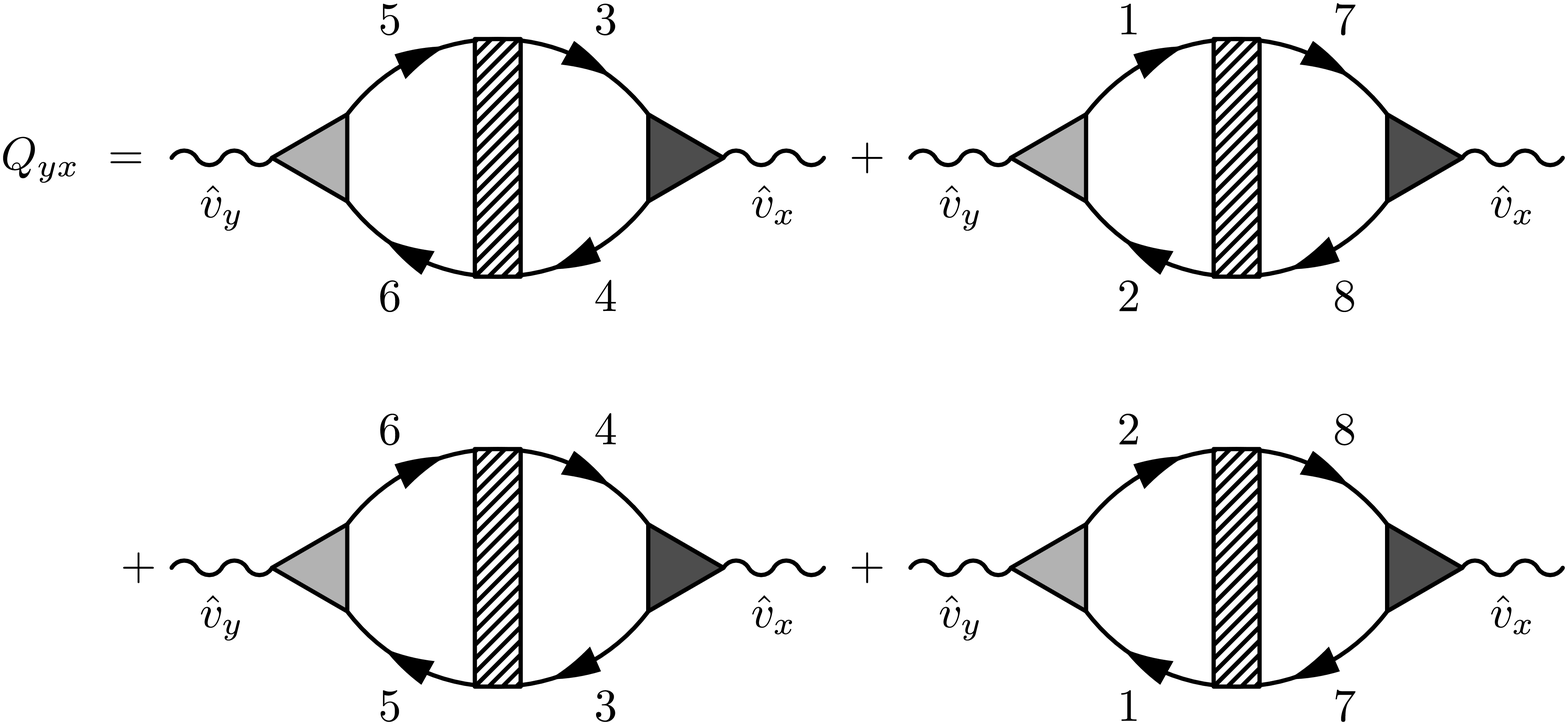}\\
\center{(a)}\vspace{0.3cm}
\includegraphics[width=\columnwidth]{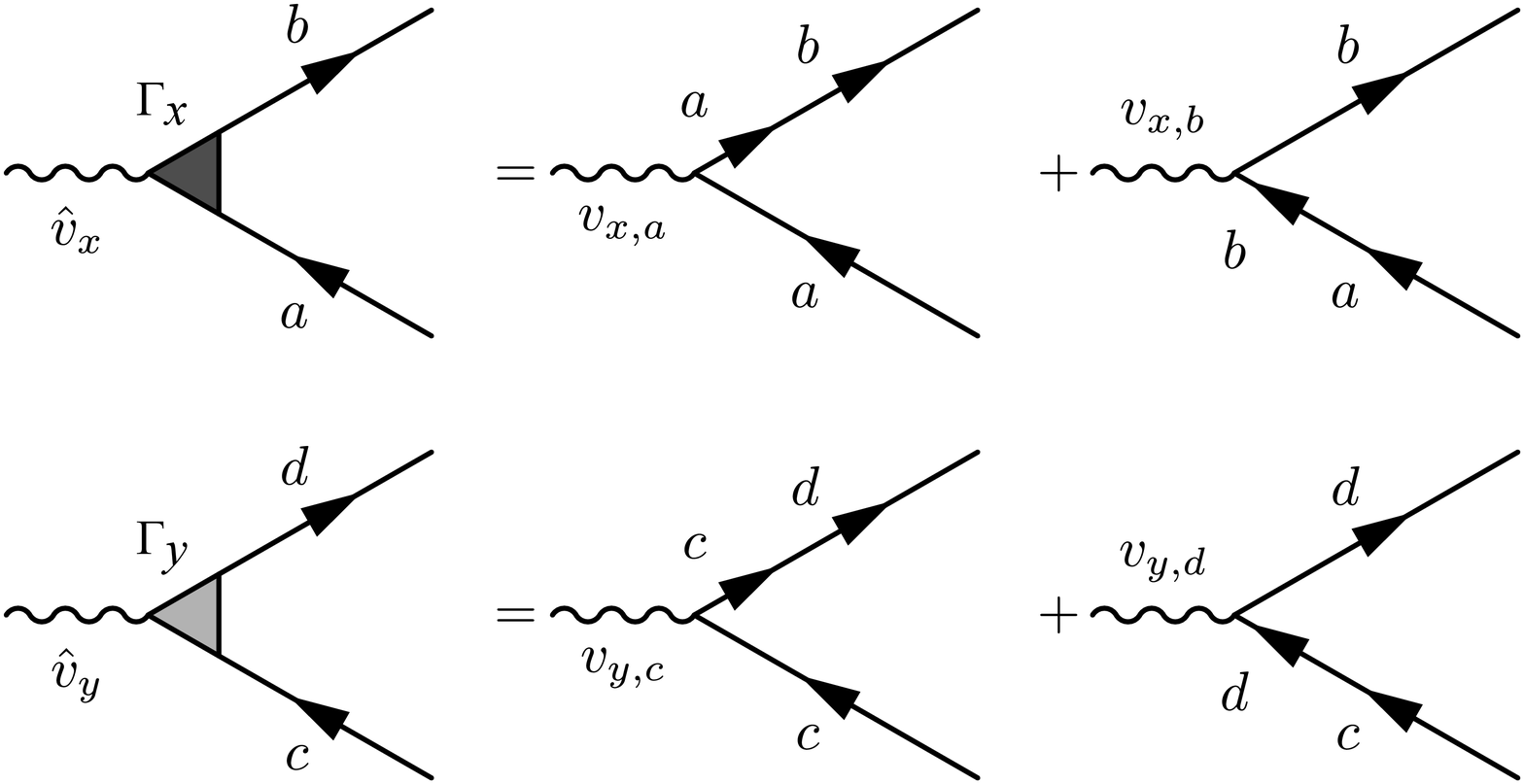}
\center{(b)}
\caption{Panel (a): Diagrammatic representations of off-diagonal current-current correlations $Q_{xy}$ and $Q_{yx}$. The shaded rectangular vertex represents impurity scattering, whose details we discuss later. Panel (b): The velocity vertices $\hat v_x$ and $\hat v_y$ convoluted with anomalous and normal Green's functions, which are labeled by $\Gamma_x$ and $\Gamma_y$. For the Fermi surface geometry we consider, $(a,b)$=(3,4) or (7,8) and $(c,d)$=(2,1) or (6,5).}
\label{Q}
\end{figure}

To simplify the calculations, we assume
 that at hot spots 1,2,5,6 the Fermi surface is vertical and  at hot spots 3,4,7,8 the Fermi surface is horizontal. In this approximation, only fermions at hot spots 3,4,7,8 couple to $\hat v_x$ vertex and only fermions at hot spots 1,2,5,6 couple to $\hat v_y$ vertex. At the end we will discuss the effect of such a Fermi surface geometry on the Kerr signal.

The diagrams
 for
 $Q_{xy}$ and $Q_{yx}$ are shown in Fig.\ 2(a). We note that in order for $Q_{xy}-Q_{yx}$ to be nonzero, both $\Delta_A$ and $\Delta_B^*$ have to be included in
   each diagram.
    Since $\Delta_A$ and $\Delta_B^*$ connect different pairs of hot spots, it follows that the Green functions on the left and right of the diagrams correspond to different pairs of hot spots. The shaded rectangular vertex must
     then transfer electrons between pairs of hot spots. The simplest way to achieve this is through impurity scattering~\cite{yakovenko, goryo}.

The filled triangles in Fig.\ 2(a) represent velocity vertices $\hat v_x$ and $\hat v_y$ convoluted with Green's functions. We label them as $\Gamma_x({\bf k}, \omega_m, \Omega_m)$ and $\Gamma_y({\bf p}, \omega_m, \Omega_m)$, as shown in Fig.\ \ref{Q}(b). Because of the simplified Fermi surface geometry, we need to  consider only $(a, b)$=(3,4) or (7,8) and $(c, d)$=(2,1) or (6,5).
In order to include both $\Delta_A$ and $\Delta_B$ into the diagram in Fig.\ \ref{Q}(a), $\Gamma_x$ and $\Gamma_y$  should each include one normal  Green's function $G({\bf k},\omega_m)$ and one ``anomalous" Green's function $F({\bf k},{\bf k+Q},\omega_m)$.
The anomalous Green's function, denoted by double arrows in Fig. \ref{Q}(b), has the form
\begin{align}
&F({\bf k-Q}/2,{\bf k+Q}/2,\omega_m)\nonumber\\
&=-\frac{\Delta_{k}^Q}{(i\omega_m-\epsilon_{k-Q}/2)(i\omega_m-\epsilon_{k+Q/2})-|\Delta_k^Q|^2}\nonumber\\
&\approx-\frac{\Delta_{k}^Q}{(i\omega_m-\epsilon_{k-Q/2})(i\omega_m-\epsilon_{k+Q/2})}.
\label{1_a}
\end{align}
where $\Delta_k^Q\equiv\langle c^\dagger_{k+Q/2}c_{k-Q/2}\rangle$. In the last step we have assumed that $\Delta_k^Q$ is much smaller than all other typical energy scales.
 Using (\ref{1_a}), we obtain for $\Gamma_x$ shown in Fig.\ \ref{Q}(b),
\begin{align}
&\Gamma_x({\bf k}_a,\omega_m,\Omega_m)\nonumber\\
=&\nonumber G({\bf k}_a,\Omega_m-\omega_m/2)\hat v_{x}F({\bf k}_a,{\bf k}_b,\Omega_m+\omega_m/2)\nonumber\\
&+F({\bf k}_a,{\bf k}_b,\Omega_m-\omega_m/2)\hat v_{x}G({\bf k}_b,\Omega_m+\omega_m/2)\nonumber\\
=&\frac{v_{x,a}\Delta_{(k_a+k_b)/2}^{k_b-k_a}}{i(\Omega_m-\omega_m/2)-\epsilon_a}\frac{1}{[i(\Omega_m+\omega_m/2)-\epsilon_b]^2}\nonumber\\
&+\frac{\Delta_{(k_a+k_b)/2}^{k_b-k_a}v_{x,b}}{[i(\Omega_m-\omega_m/2)-\epsilon_a]^2}\frac{1}{i(\Omega_m+\omega_m/2)-\epsilon_b}\nonumber\\
=&\frac{2(i\Omega_m-\epsilon_a)~v_F\Delta_{(k_a+k_b)/2}^{k_b-k_a}}{[i(\Omega_m-\omega_m/2)-\epsilon_a]^2[i(\Omega_m+\omega_m/2)-\epsilon_a]^2},
\end{align}
where in the last line we have used the fact that
 in the approximation we are using, the Fermi surfaces for $(a,b)=$(3,4) or (7,8)
 are parallel, hence $v_{x,a}=v_{x,b}=v_F$ and $\epsilon_a=\epsilon_b$.
  Similarly,
\begin{align}
&\Gamma_y({\bf p}_c,\omega_m,\Omega_m)\nonumber\\
=&\nonumber G({\bf p}_c,\Omega_m-\omega_m/2)\hat v_{y}F({\bf p}_c,{\bf p}_d,\Omega_m+\omega_m/2)\nonumber\\
&+F({\bf p}_c,{\bf p}_d,\Omega_m-\omega_m/2)\hat v_{y}G({\bf p}_d,\Omega_m+\omega_m/2)\nonumber\\
=&\frac{v_{y,c}\Delta_{(p_c+p_d)/2}^{p_d-p_c}}{i(\Omega_m-\omega_m/2)-\epsilon_c}\frac{1}{[i(\Omega_m+\omega_m/2)+\epsilon_d]^2}\nonumber\\
&+\frac{\Delta_{(p_c+p_d)/2}^{p_d-p_c}v_{y,d}}{[i(\Omega_m-\omega_m/2)-\epsilon_c]^2}\frac{1}{i(\Omega_m+\omega_m/2)-\epsilon_d}\nonumber\\
=&\frac{(i\omega_m+2\epsilon_c)~v_F\Delta_{(p_c+p_d)/2}^{p_d-p_c}}{[(\Omega_m-\omega_m/2)^2+\epsilon_c^2][(\Omega_m+\omega_m/2)^2+\epsilon_c^2]},
\end{align}
where  in the last line we have used the fact that for $(c,d)=$(2,1) or (6,5), Fermi surfaces are anti-parallel, thus $v_{y,c}=-v_{y,d}=v_F$ and $\epsilon_c=\epsilon_d$.

The eight diagrams
 in Fig.\ 2(a) are related by symmetry.
 First,
    it is trivial to identify that the diagrams in the left column are related to the ones in the right column by mirror reflection in $x$ direction and that the diagrams in the upper row are related to the ones in the lower row by mirror reflection in $y$
direction. Second, $Q_{xy}$ and $Q_{yx}$ are related by time reversal. This point can be easily seen diagrammatically: by reversing the momentum and arrows $Q_{xy}$ transforms into $Q_{yx}$.
If time reversal symmetry or mirror symmetries are not broken, then these diagrams will cancel out each other. Since our order parameter $\Upsilon$ is odd under all these transformations, all of them will add up constructively.
 Using these arguments, we need to only evaluate one diagram and
  add others using symmetry.

  Evaluating the first diagram in Fig.\ 2(a),
   we obtain that in Matsubara frequencies
\begin{align}
Q&_{xy}(\omega_m)=-4e^2\langle\Delta_A^*\Delta_B\rangle  v_F^2 \nonumber\\
&\times 2T\sum_{\Omega_m}V(\omega_m,\Omega_m)\tilde\Gamma_x(\omega_m,\Omega_m)\tilde\Gamma_y(\omega_m,\Omega_m),
\label{27}
\end{align}
and
\begin{align}
Q&_{yx}(\omega_m)=-4e^2\langle\Delta_B^*\Delta_A\rangle  v_F^2 \nonumber\\
&\times 2T\sum_{\Omega_m}V(\omega_m,\Omega_m)\tilde\Gamma_x(\omega_m,\Omega_m)\tilde\Gamma_y(\omega_m,\Omega_m),
\end{align}
where the factor 4
 counts the number of diagrams for each correlator, the factor 2 comes from summing over spin indices,
  $V(\omega_m,\Omega_m)$ in Eq.\ (\ref{27}) is impurity scattering potential which does not depend on momentum, and
\begin{widetext}
\begin{align}
\tilde\Gamma_x(\omega_m,\Omega_m)=&\int_{-\Lambda}^{\Lambda}\frac{dk_xdk_y}{(2\pi)^2}\frac{2(i\Omega_m+v_F k_x)}{[i(\Omega_m-\omega_m/2)+v_Fk_x]^2[i(\Omega_m+\omega_m/2)+v_Fk_x]^2}\nonumber\\
\tilde\Gamma_y(\omega_m,\Omega_m)=&\int_{-\Lambda}^{\Lambda}\frac{dp_xdp_y}{(2\pi)^2}\frac{i\omega_m+2v_Fp_y}{[(\Omega_m-\omega_m/2)^2+v_F^2p_y^2][(\Omega_m+\omega_m/2)^2+v_F^2p_y^2]}.
\label{ywfri2}
\end{align}
\end{widetext}
In (\ref{ywfri2}) we linearized the dispersion relation in the vicinity of hot spots and  used the fact that
$\Delta_{k_0}^{-Q}=\(\Delta_{k_0}^Q\)^*=\Delta_A^*$ and $\Delta_{{k_0+(\pi,\pi)}}^Q=-\Delta_B$.

We point out that
 $Q_{xy}$ is not
   affected
  by the $d$-form factor of the CDW order parameter because the CDW order parameters appear in bilinear form
and both
change sign simultaneously
under the shift by $(\pi,\pi)$.
 We emphasize that this property is specific to the Fermi surface geometry which we consider.

Using the fact that $\langle\Delta_A^*\Delta_B-\Delta_A\Delta_B^*\rangle=i\Upsilon$, we find that
\begin{align}
&Q_{xy}(\omega_m)-Q_{yx}(\omega_m)\nonumber\\
=&-8ie^2\Upsilon  v_F^2 T\sum_{\Omega_m}V(\omega_m,\Omega_m)\tilde\Gamma_x(\omega_m,\Omega_m)\tilde\Gamma_y(\omega_m,\Omega_m).
\label{yw3}
\end{align}
We see from Eq.\ (\ref{yw3}) that the
Kerr angle is non-zero
 if ${\tilde \Gamma}_x$ and ${\tilde \Gamma}_y$ are non-zero. This, however, is not always the case.
In particular, ${\tilde \Gamma}_x$ vanishes
 if the integration  over $k_x$ in (\ref{ywfri2}) is over the whole momentum range, from $-\infty$ to $\infty$. To see this, we calculate the residues of the integrand at two poles in $k_x$,
 at $k_x = i(\Omega_m\pm\omega_m/2)/v_F$. Both
 are double poles. Denoting the
 integrand as $I$ and evaluating the residues at $i(\Omega_m\pm\omega_m/2)/v_F$, we obtain
\begin{align}
&{\rm Res}(I,i(\Omega_m+\omega_m/2)/v_F)\nonumber\\
=&\frac{d}{dk_x}\left\{\frac{2i\Omega_m+2v_Fk_x}{[i(\Omega_m-\omega_m/2)+v_Fk_x]^2}\right\}\bigg|_{k_x=i(\Omega_m+\frac{\omega_m}2)/v_F}\nonumber\\
=&-2v_F\frac{i(\Omega_m+\omega_m/2)-v_Fk_x}{\[i(\Omega_m-\omega_m/2)+v_Fk_x\]^3}\bigg|_{k_x=i(\Omega_m+\frac{\omega_m}2)/v_F}\nonumber\\
=&0,\\\nonumber\\
&{\rm Res}(I,i(\Omega_m-\omega_m/2)/v_F)\nonumber\\
=&\frac{d}{dk_x}\left\{\frac{2i\Omega_m+2v_Fk_x}{[i(\Omega_m-\omega_m/2)+v_Fk_x]^2}\right\}\bigg|_{k_x=i(\Omega_m-\frac{\omega_m}2)/v_F}\nonumber\\
=&-2v_F\frac{i(\Omega_m-\omega_m/2)-v_Fk_x}{\[i(\Omega_m+\omega_m/2)+v_Fk_x\]^3}\bigg|_{k_x=i(\Omega_m-\frac{\omega_m}2)/v_F}\nonumber\\
=&0.
\end{align}
As a result, ${\tilde \Gamma}_x=0$.

This vanishing of ${\tilde \Gamma}_x$ is, however, an artifact because ${\bf k}$ is a deviation from a hot spot and the linearization of the fermionic dispersion near a hot spot is only valid in a finite window of ${\bf k}$. To account for this, we introduce a finite momentum cutoff $\Lambda$ (same for $k_x$ and $k_y$).
Evaluating $\Gamma_x$ and $\Gamma_y$ by integrating over $k_x$ and $k_y$
 $-\Lambda$ to $\Lambda$, we obtain,
\begin{align}
&\tilde\Gamma_x(\omega_m,\Omega_m)\nonumber\\
&=\frac{2i\Lambda^2\Omega_m}{\pi^2[(\Omega_m+\omega_m/2)^2+v_F^2\Lambda^2][(\Omega_m-\omega_m/2)^2+v_F^2\Lambda^2]}\label{yw222}
\end{align}
and
\begin{align}
&\tilde\Gamma_y(\omega_m,\Omega_m)\nonumber\\
&=\frac{i\Lambda}{4\pi v_F\Omega_m}\[\frac{1}{|\Omega_m+\omega_m/2|}-\frac{1}{|\Omega_m-\omega_m/2|}\].
\label{yw2}
\end{align}
  Now ${\tilde \Gamma}_x$  is finite.  We note, however, that ${\tilde \Gamma}_x$ is odd in running frequency $\Omega_m$ while ${\tilde \Gamma}_y$ is even in
   $\Omega_m$. Then, when we substitute ${\tilde \Gamma}_x (\omega_m,\Omega_m)$ and ${\tilde \Gamma}_y (\omega_m,\Omega_m)$ into Eq. (\ref{27}) for ${Q}_{xy} (\omega_m)$ and sum over the running frequency $\Omega_m$, we find that ${Q}_{xy} (\omega_m)$ vanishes for any $V(\omega_m, \Omega_m)$, which is even in $\Omega_m$. In particular, ${Q}_{xy} (\omega_m)$ vanishes for
a Gaussian-type disorder scattering, for which $V(\omega_m, \Omega_m)$ does not depend on $\Omega_m$.

A similar problem emerges in the analysis of Kerr effect in
SrRuO$_4$, (Refs.~\cite{goryo,yakovenko,mineev,kallin}), which is believed by many to be
a $p+ip$ superconductor (Refs. \cite{ueda,maeno,andy,rice_s})
 We follow the same strategy as these authors used to obtain a non-zero ${Q}_{xy} (\omega_m)$ and (i) consider
skew scattering from non-Gaussian disorder, for which $V(\omega_m,\Omega_m)$ is odd in $\Omega_m$, and (ii)
include particle-hole asymmetry, which gives rise to the appearance of an even in $\Omega_m$ term $\tilde\Gamma_x\tilde\Gamma_y$.\\

\subsubsection{Skew scattering}
\begin{figure}[h]
\includegraphics[width=\columnwidth]{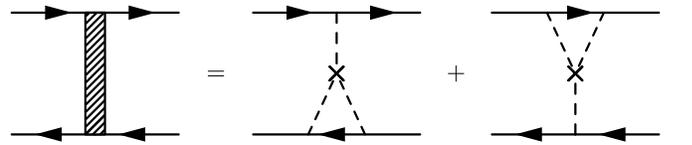}
\caption{The skew scattering vertex.}
\label{skew}
\end{figure}

First we consider the case of skew scattering. The scattering vertex is characterized by a third cumulant (skewness) of impurity potentials,
\begin{align}
\langle V_{\rm imp}(q_1)V_{\rm imp}(q_2)&V_{\rm imp}(q_3)\rangle\nonumber\\
&=\kappa_3n_iu_0^3\delta(q_1+q_2+q_3).
\end{align}
Here $n_i$ is the impurity concentration, and $\kappa_3$ is the dimensionless parameter characterizing the skewness, which varies from 0 to 1 depending on the deviation of the actual distribution function from the Gaussian. $\kappa_3=1$ corresponds to
 the case when all impurities scatter with
 with equal strength $u_0$, and $\kappa_3=0$ corresponds to
  the case when each impurity scatters with equal probability for
  attraction and repulsion.
We replace the shaded impurity scattering vertex with skew scattering vertex, shown in Fig.\ \ref{skew}.

Evaluating the diagrams in Fig. 3 we
 obtain for the scattering potential,
\begin{widetext}
\begin{align}
V^{(a)}(\omega_m,\Omega_m)=&\kappa_3n_iu_0^3\int\frac{d^2k}{(2\pi)^2}\[G(\Omega_m+\omega_m/2,k)+G(\Omega_m-\omega_m/2,k)\]\nonumber\\
\approx
&i\pi\kappa_3n_iu_0^3N(0)[\sgn(\Omega_m+\omega_m/2)+\sgn(\Omega_m-\omega_m/2)]
\label{yw1}
\end{align}
where
$N(0)$ is the density of states averaged over the Fermi surface
\begin{align}
N(0)=\int\frac{d\theta}{2\pi}N(0,\theta).
\end{align}

We note here that $V^{(a)}$ is {\it odd} in frequency $\Omega_m$. With $\tilde\Gamma_{x}\tilde\Gamma_{y}$ also being odd in $\Omega_m$, now
$\tilde\Gamma_{x}\tilde\Gamma_{y} V^{(a)}$ is even in $\Omega_m$ and the result of summation over $\Omega_m$ in
Eq.\ (\ref{yw3})
 is  non-zero. Plugging  Eqs. (\ref{yw222},\ref{yw2},\ref{yw1}) into Eq. (\ref{yw3}) we find that
\begin{align}
Q_{xy}^{(a)}(\omega_m)-Q_{yx}^{(a)}(\omega_m)=&\frac{-4e^2}{\pi^2}\Upsilon v_F\kappa_3n_iu_0^3N(0)\Lambda^3\nonumber\\
&\times T\sum_{\Omega_m}\frac{\sgn(\Omega_m+\omega_m/2)+\sgn(\Omega_m-\omega_m/2)}{[(\Omega_m+\omega_m/2)^2+v_F^2\Lambda^2][(\Omega_m-\omega_m/2)^2+v_F^2\Lambda^2]}\[\frac{1}{|\Omega_m+\omega_m/2|}-\frac{1}{|\Omega_m-\omega_m/2|}\]\nonumber\\
=&\frac{16e^2}{\pi^2}\Upsilon v_F\kappa_3n_iu_0^3N(0)\Lambda^3\nonumber\\
&\times T\sum_{\Omega_m>|\omega_m|/2}
\frac{1}{[(\Omega_m+\omega_m/2)^2+v_F^2\Lambda^2][(\Omega_m-\omega_m/2)^2+v_F^2\Lambda^2]}\frac{\omega_m}{\Omega_m^2-\omega_m^2/4}.
\label{37}
\end{align}
\end{widetext}

In the
 range  $T\ll\omega_m\ll v_F\Lambda$,
  relevant to experiments,
  we find that the
   frequency sum is infra-red divergent, when replaced by the integral.  The divergence is cut by $T$ and the
    result is
\begin{align}
Q_{xy}^{(a)}(\omega_m)-Q_{yx}^{(a)}(\omega_m)=&\frac{8e^2\Upsilon\kappa_3n_iu_0^3N(0)}{\pi^3v_F^3\Lambda}\nonumber\\
&\times\sgn(\omega_m)\log\frac{|\omega_m|}{T}.
\end{align}
After analytical continuation $\omega_m\to-i\omega+\delta$, we find the retarded current-current correlator
\begin{align}
Q_{xy}^{(a),R}(\omega)-Q_{yx}^{(a),R}(\omega)=&\frac{8e^2\Upsilon\kappa_3n_iu_0^3N(0)}{\pi^3v_F^3\Lambda}\nonumber\\
&\times\[\log\frac{|\omega|}{T}-i\frac{\pi}{2}\sgn(\omega)\].
\label{yw5}
\end{align}
Plugging this back into Eqs.\ ({\ref{9},\ref{kubo}}), we find the skew-scattering contribution to the Kerr angle to be
\begin{align}
\theta_K^{(a)}=\frac{-16N(0)e^2\Upsilon\kappa_3n_iu_0^3}{\pi^2n(n^2-1)v_F^3\Lambda d}\frac{1}{\omega^2}\log\frac{|\omega|}{T}.
\label{ywfri1}
\end{align}
Eq.\ (\ref{ywfri1}) is the first main result of this paper:
 $\theta_K$ is proportional to $\Upsilon$ and the prefactor depends on frequency as $(1/\omega^2) \log{|\omega|/T}$.
   We caution that this result is valid in the limit $\omega\gg T$, and
 cannot
  be directly extrapolated to the DC limit $\omega\to0$. 
  Also, in deriving Eq. \ (\ref{ywfri1}) we linearized fermionic dispersion near the Fermi surface. This formula is then valid only for $\omega$ smaller than the 
   scale at which the curvature of the Fermi surface becomes relevant.

\subsubsection{Gaussian-type disorder with particle-hole asymmetry}

Next we consider the contribution
to ${Q}_{xy}$ from  Gaussian-type disorder with
\begin{align}
\langle V_{\rm imp}(q_1)V_{\rm imp}(q_2)\rangle=n_iu_0^2\delta^2(q_1+q_2).
\label{ywsat1}
\end{align} \begin{figure}
\includegraphics[width=0.633\columnwidth]{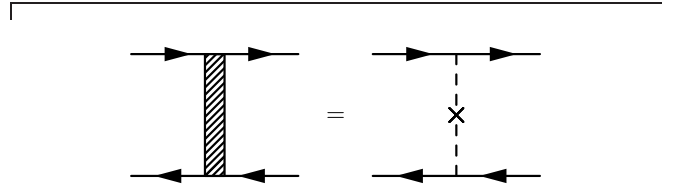}
\caption{The scattering vertex from a Gaussian-type impurity.}
\label{gauss}
\end{figure}
 The impurity potential is shown in Fig.\ \ref{gauss}.  For this potential
$V^{(b)}=n_iu_0^2$.
Like we said, for particle-hole symmetric case $Q_{xy}$ for such $V^{(b)}$ vanishes because
$V^{(b)}$ is independent on $\Omega_m$ while $\tilde\Gamma_{x}\tilde\Gamma_{y}$ is odd in $\Omega$.
In order to obtain an non-zero result, particle-hole asymmetry in Eq.\ (\ref{ywfri2}) has to be included.
The simplest way to introduce particle-hole asymmetry is to
 change the integration range over $k_x$ in  Eq. (\ref{ywfri2})  from  $-\Lambda$ to $\Lambda$ to  $-\Lambda$ to $\Lambda+\delta\Lambda$.
  A non-zero $\delta \Lambda$ gives rise to
  a term  in ${\tilde  \Gamma}_x$ which is
  even in $\Omega_m$.
  At the same time,
  $\tilde\Gamma_y$ given by Eq.\ (\ref{ywfri2})
  is  even in $\Omega_m$ independent on whether or
    not we include particle-hole asymmetry.
    Combining $\Omega$-symmetric ${\tilde \Gamma}_y V^{(b)}$ with $\Omega$-symmetric piece in ${\tilde \Gamma}_x$ we obtain a non-zero contribution to
     $Q_{xy}$.

To estimate the magnitude of this contribution we assume that the particle-hole asymmetry is small, namely $\delta\Lambda\ll\Lambda$.
Then
\begin{align}
\tilde\Gamma_{x}(\omega_m,\Omega_m)\to\tilde\Gamma_{x}(\omega_m,\Omega_m)+\delta\tilde\Gamma_{x}(\omega_m,\Omega_m),
\label{40}
\end{align}
where,
\begin{align}
\delta\tilde\Gamma_x=&\frac{\Lambda}{2\pi^2}\int_{\Lambda}^{\Lambda+\delta\Lambda}dk_x\nonumber\\
&\times\frac{2i\Omega_m+2v_Fk_F}{\[i(\Omega_m-\omega_m/2)+v_Fk_x\]^2\[i(\Omega_m+\omega_m)+v_Fk_x\]^2}\nonumber\\
\approx&\frac{\delta\Lambda}{\pi^2v_F^3\Lambda^2}.
\label{ywsat2}
\end{align}
In the last step we
 that $\Omega_m,~\omega_m,~v_F\delta\Lambda\ll v_F\Lambda$.
 Since $\delta\tilde\Gamma_{x}(\omega_m,\Omega_m)$ is already small,
 it is sufficient to use the
 result for $\tilde\Gamma_y$ from particle-hole symmetric case
\begin{align}
\tilde\Gamma_y
=\frac{i\Lambda}{4\pi v_F\Omega_m}\[\frac{1}{|\Omega_m+\omega_m/2|}-\frac{1}{|\Omega_m-\omega_m/2|}\].
\label{aug151}
\end{align}
From Eqs.\ (\ref{yw3},\ref{40},\ref{ywsat2},\ref{aug151}) we obtain for the current-current correlator
\begin{align}
Q_{xy}^{(b)}(\omega_m)-Q_{yx}^{(b)}(\omega_m)=&-\frac{2e^2\Upsilon n_iu_0^2{\delta\Lambda}}{\pi^3v_F\Lambda} \gamma(\omega_m),
\label{43}
\end{align}
where
\begin{align}
\gamma(\omega_m)=&T\sum_{\Omega_m}\frac{1}{\Omega_m}\(\frac{1}{|\Omega_m+\omega_m/2|}-\frac{1}{|\Omega_m-\omega_m/2|}\)\nonumber\\
=&T\sum_{\Omega_m'}\frac{1}{|\Omega_m'|}\frac{\omega_m}{(\Omega_m')^2-\omega_m^2/4}.
\label{ywsun5}
\end{align}
In the last step we have shifted the running frequency $\Omega_m$ to
$\Omega_m'=\Omega_m\pm\omega_m/2$.
 Evaluating the sum and cutting the logarithmical divergence by $T$
  we obtain that in the limit $T\ll\omega_m$,
\begin{align}
\gamma(\omega_m)=-\frac{4}{\pi\omega_m}\log\frac{|\omega_m|}{T}.
\label{ywsun1}
\end{align}
In Appendix \ref{app:1}, we present an alternative derivation of Eq.\ (\ref{ywsun1}).

Substituting Eq. (\ref{ywsun1}) into  Eq. (\ref{43}) we
obtain
\begin{align}
Q_{xy}^{(b)}(\omega_m)-Q_{yx}^{(b)}(\omega_m)=&\frac{8e^2\Upsilon n_iu_0^2{\delta\Lambda}}{\pi^4v_F^2\Lambda\,\omega_m}\log\frac{|\omega_m|}{T}.
\end{align}
After the analytical continuation $\omega_m\to-i\omega+\delta$ we find the retarded current-current correlator
\begin{align}
Q_{xy}^{(b),R}(\omega)-Q_{yx}^{(b),R}(\omega)=&\frac{8ie^2\Upsilon n_iu_0^2{\delta\Lambda}}{\pi^4v_F^2\Lambda\,\omega}\nonumber\\
&\times\[\log\frac{|\omega|}{T}-i\frac{\pi}{2}\sgn(\omega)\].
\end{align}
Plugging this
 into Eqs.\ ({\ref{9},\ref{kubo}}) we find the contribution to the Kerr angle from
 particle-hole asymmetry  to be
\begin{align}
\theta_K^{(b)}=\frac{-8e^2\Upsilon n_iu_0^2\delta\Lambda}{\pi^2n(n^2-1)v_F^2\Lambda d}\frac{1}{|\omega|^3}.
\label{ywsun2}
\end{align}
Eq.\ (\ref{ywsun2}) is the second main result of this paper. Once again, we find that  the Kerr angle
 is proportional to $\Upsilon$ and the proportionality coefficient is a decreasing function of $\omega$, this time
 $1/|\omega|^3$.
 The  ratio of the contributions from particle-hole asymmetry and skew scattering is, roughly,
  $\sim v_F\delta\Lambda/(\kappa_3\omega)$. 
  Like before, this formula  valid only for $\omega$ smaller than the
   scale at which the curvature of the Fermi surface becomes relevant.

\subsection{Kerr effect due to spin-fluctuations}
\label{spf}
\begin{figure}[htp]
\includegraphics[width=0.633\columnwidth]{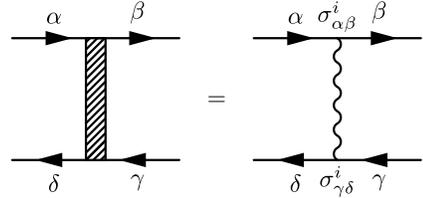}
\caption{The scattering vertex from the antiferromagnetic spin fluctuation. A Pauli matrix $\sigma^i$ is associated to each vertex.}
\label{sf}
\end{figure}

 As we already said,
 in order to get a
 non-zero
 Kerr effect, we need a vertex (shaded rectangle in Fig.\ 2) which
 connects  hot fermions from pairs with center of mass momentum $k_0$ and $-k_0$,  say fermions with momenta near points 1 and 7 in in Fig.\ 1,
  otherwise the combination $\Delta^*_A\Delta_B - \Delta^*_B \Delta_A$ would not emerge. We remind that $k_0 = (\pi -Q/2,0)$, where $Q$ is the momentum of incommensurate CDW order.
 Impurity scattering can accomplish this -- but so can antiferromagnetic spin fluctuations
 with momentum transfer
  near
  $(\pi \pm Q, \pi)$.
   In this subsection we discuss the Kerr effect from exchange of spin fluctuations in a clean (disorder free) system.
   The analysis parallels the one for
    impurity scattering, and
     the only change we need to make is to replace the impurity-mediated interaction in Fig.\ \ref{Q} with
      the one mediated by spin fluctuations, see
       Fig.\ \ref{sf}.

 In the vicinity of the quantum-critical point of anti-ferromagnetism  the interaction mediated by spin fluctuations depends strongly on both momentum and frequency transfer. The propagator of spin fluctuation can be written as~\cite{acs}
\begin{align}
\chi_{\alpha\beta,\gamma\delta}({\bf q},\omega_n)=\frac{\bar g\ \sigma^i_{\alpha\beta}\sigma^i_{\gamma\delta}}{({\bf q-K})^2+\gamma|\omega_n|+\xi^{-2}},
\end{align}
where ${\bf K}=(\pi,\pi)$, $\bar g$ describes the coupling between spin fluctuations and fermions, $\sigma^i_{\alpha\beta}$ is the Pauli matrix, $\bf q$ and $\omega_n$ are respectively momentum and frequency transfer, $\xi$ is the magnetic correlation length, and $\gamma|\omega_n|$ with $\gamma=4\bar g/(\pi v_F^2)$
 describes the Landau damping of a spin fluctuation.

First, we show that $Q_{xy}$ vanishes if we assume particle-hole symmetry, i.e., linearize fermionic dispersion near hot spots and intergate over momentum deviation from a hot spot in symmetric limits.
As before, symmetry arguments allow us
to restrict with the evaluation of only the first diagram in Fig.\ \ref{Q}. In this diagram, a spin fluctuation propagator has
 to connect fermions at hot spot pairs (1,2) and (7,8). Defining the fermionic momenta with respect to the position of each hot spots, we find that
\begin{align}
&\chi_{\alpha\beta,\gamma\delta}({\bf k-p},\Omega_m-\Omega_m')\nonumber\\
&=\frac{\bar g\ \sigma^i_{\alpha\beta}\sigma^i_{\gamma\delta}}{(k_y-p_y)^2+(k_x-p_x-Q)^2+\gamma|\Omega_m-\Omega_m'|+\xi^{-2}},
\end{align}
Following the same procedures that led us to Eq.\ (\ref{37}), we obtain for the current-current correlator
\begin{align}
&Q_{xy}^{(c)}(\omega_m)-Q_{yx}^{(c)}(\omega_m)\nonumber\\
=&-4ie^2\Upsilon v_F^2 T^2\sum_{m,m'}\int\frac{d^2kd^2p}{(2\pi)^4}\nonumber\\
&\times\frac{2(i\Omega_m+v_Fk_x)}{[i(\Omega_m-\omega_m/2)+v_Fk_x]^2[i(\Omega_m+\omega_m/2)+v_Fk_x]^2}\nonumber\\
&\times\frac{i\omega_m+2v_Fp_y}{[(\Omega_m'-\omega_m/2)^2+v_F^2p_y^2][(\Omega_m'+\omega_m/2)^2+v_Fp_y^2]}\nonumber\\
&\times\frac{6\bar g}{(k_y-p_y)^2+(k_x-p_x-2Q)^2+\gamma|\Omega_m-\Omega_m'|+\xi^{-2}},
\end{align}
where the factor 6 in the last line comes from summation over spin indices:  $\Tr(\sigma^i_{\alpha\beta}\delta_{\beta\gamma}\sigma^i_{\gamma\delta}\delta_{\delta\alpha})=6$.
The key observation here is that, because of the Fermi surface geometry and particle-hole symmetry, integration parameters $p_x$ and $k_y$ only appear in boson propagator (the last line).
Shifting the variables we obtain
\begin{align}
Q_{xy}^{(c)}&(\omega_m)-Q_{yx}^{(c)}(\omega_m)
=-4ie^2\Upsilon  v_F^2 \nonumber\\
&\times T^2\sum_{m,m'}\bar\chi(\Omega_m-\Omega_m')\tilde\Gamma_x(\omega_m,\Omega_m)\tilde\Gamma_y(\omega_m,\Omega_m'),
\label{sfs}
\end{align}
where $\tilde\Gamma_x$ and $\tilde\Gamma_y$ are
the same as in Eq.\ (\ref{ywfri2}) and
\begin{align}
\bar\chi(\Omega_m-\Omega_m')=\frac{3}{2\Lambda^2}\int\frac{dXdY~\bar g}{X^2+Y^2+\gamma|\Omega_m-\Omega_m'|+\xi^{-2}}.
\end{align}
 Like in the case of impurity scattering,
  $\tilde\Gamma_x$ is odd in $\Omega_m$ and $\tilde\Gamma_y$ is even in $\Omega_m'$. Since $\bar\chi$ is even in $\Omega_m-\Omega_m'$, the integral
  vanishes.

 In order to obtain non-zero Kerr signal we then
    need to include
    particle-hole asymmetry in the fermion dispersion, like
     we did in the case of Gaussian-type disorder. Once we make the limits of momentum integration asymmetric,
       $\tilde\Gamma_x$ acquires an even component $\delta \tilde\Gamma_x$, and the
       combination  $\delta \tilde\Gamma_x \tilde\Gamma_y \bar\chi$ is even in frequency and does not vanish after the summations over $\Omega_m$ and $\Omega'_m$ in (\ref{sfs}).
   Following the same steps as in the impurity case we obtain
         for the current-current correlator,
\begin{align}
Q_{xy}^{(c)}(\omega_m)-Q_{yx}^{(c)}(\omega_m)=&-\frac{2e^2\Upsilon\lambda{\delta\Lambda}}{\pi^3v_F\Lambda} \gamma(\omega_m).
\label{54}
\end{align}
The only difference between Eq.\ (\ref{54}) and Eq.\ (\ref{43}) is that we have replaced the impurity potential $n_iu_0^2$ with
\begin{align}
\lambda=\frac{3T}{4\Lambda^2}\sum_{\omega_n}^{\Omega_0}\int^{\Lambda_0}\frac{dXdY~\bar g}{X^2+Y^2+\gamma|\omega_n|+\xi^{-2}},
\end{align}
where $\Omega_0$ and $\Lambda_0$ are the upper limits of frequency summation and momentum integration.
 For the Kerr angle we then obtain
\begin{align}
\theta_K^{(c)}=\frac{-8e^2\Upsilon \lambda\delta\Lambda}{\pi^2n(n^2-1)v_F^2\Lambda d}\frac{1}{|\omega|^3},
\label{56}
\end{align} for $\omega\gg T$.
Eq. (\ref{56}) is the third main result of this work.

\subsection{Effect of the Fermi Surface geometry}
\label{fsg}
 In the calculations of $Q_{xy}$ in the preceding Sections we assumed, motivated by the actual Fermi surface geometry in the cuprates,
 that only one component of the Fermi velocity, either  $v_x$ or $v_y$, is non-zero at any given hot spot.
    Let's now consider the
     opposite limit in which
       at
        any
         hot spot $|v_x|=|v_y|$, i.e., for each hot fermion the current vertex contains
           $v_x$ and $v_y$
            components with
             equal magnitudes.
                Under this condition, the local Fermi surface geometry will be exactly the same for hot spots pairs (5,6) and (7,8), and for (3,4) and (1,2). On the other hand, due to the $d$-form factor, the CDW order parameter changes sign between the regions (5,6) and (7,8), and between the regions (3,4) and (1,2).
 Then,
 if one considers interaction mediated by
 impurity scattering, which is independent on a momentum transfer, the contribution to $Q_{xy}$ from hot regions (12,78) and (34,56)
  will be
  canceled out by that from the regions (34,78) and (12,56).
   It is easy to verify
   that all other contributions to $Q_{xy}$ also cancel out.
    Then the Kerr angle $\theta_K$ vanishes
   even though all the symmetry requirements are satisfied by the order parameter of the system.
    This shows that the Fermi surface geometry is very important for the Kerr effect.
     It is straightforward to show that in the limit that we considered
    (only one component of the Fermi velocity is non-zero at any given hot spot) $\theta_K$ is at maximum.  Once Fermi surface  geometry changes and
     both $v_x$ and $v_y$ become non-zero at any given hot spot, $\theta_K$ drops and eventually vanishes when at any hot spot $|v_x|=|v_y|$.

     The spin-fluctuation contribution to $\theta_K$ does not completely vanish when at any spot $|v_x|=|v_y|$
      because the interactions depends on momentum transfer. However,
       it is still much reduced compared to the case when only $v_x$ or $v_y$ is non-zero at any hot spot.
In this sense, the geometry of the Fermi surface in the cuprates, with near-perfect nesting in hot regions, is the ``best case"
 scenario for the Kerr effect.

\section{Effect of applying a uniform field and flipping the sample}
\label{disc}

  Finally, we briefly discuss the effect of an applied magnetic field on the Kerr effect.
    The composite order $\Upsilon$
    produces charge density and current density modulations which are spatially correlated~\cite{charge,tsvelik}, although at any given bond
     charge density and current density modulations, taken separately, vanish due to unbroken $U(1)$ phase symmetry.
     The current density modulations produces a magnetic field. However, this magnetic field is oscillatory in space,
     $H(r) \propto \cos{{\bf Q} {\bf r}}$, hence it does not couple directly to a uniform magnetic field. As a result, the sign of $\theta_K \propto \Upsilon$
      does not necessary flip when one flips the sign of the uniform magnetic field. In other words, the
    Kerr effect
    in our
    case
     is not  trained by an external magnetic field.

On the other hand, for any two-dimensional system, ``flipping the sample" amounts to a mirror reflection with respect to a vertical plane.
  The Kerr angle should necessarily reverse the sign under such transformation (see Table \ref{tab1}).
  To explain the absence of such a sign change, one likely has to
   analyze
   the three-dimensional structure of the system~\cite{yakovenko_future},
   since in three dimensions ``flipping the sample"  also introduces $z\to-z$. Such 3D analysis is, however,  beyond the scope of this paper.

\section{Conclusions}
\label{conc}

In this work, we have studied the
Kerr angle $\theta_K$ in a state with preemptive charge order which breaks $C_4$ lattice rotational symmetry down to $C_2$ and breaks $Z_2$ time-reversal symmetry, but preserves $U(1)$ phase symmetry which gets broke only for a true CDW-ordered state. We called such state chiral-nematic. We argued that
the Kerr angle is
 related to the antisymmetric component of the Hall conductivity $\sigma_{xy}-\sigma_{yx}$, even when the $C_4$ lattice rotation symmetry is broken. We derived the expression for $\theta_K$ for a system with incommensurate CDW order, and demonstrated that the
  result only depends on the chiral-nematic charge order parameter $\Upsilon$.
   Therefore, even if the CDW order is not yet developed, a preemptive composite charge order already
    gives rise to a non-zero $\theta_K$.
 We have shown that a non-zero $\theta_K \propto \Upsilon$ is allowed by the symmetry of chiral-nematic order, however microscopic analysis is needed to
 determine the prefactor and verify that it is not equal to zero. We demonstarted that
 a non-zero $\theta_k$ emerges due to
  either impurity scattering or scattering by spin fluctuations, and one additionally needs either skew scattering from the impurities or particle-hole asymmetry. Since the chiral-nematic order
  produces spatially oscillatory magnetic fields,
   the Kerr signal in our case
   cannot be trained by a uniform
    magnetic
     field.

\begin{acknowledgments}
We are grateful to E. Berg, I. Eremin, A. Maharaj, S. Kivelson, S. Lederer, L. Nie, J. Orenstein, S. Raghu, V. Yakovenko and particularly A. Kapitulnik
for fruitful discussions.
This work was supported by the DOE grant DE-FG02-ER46900 (A.C. and Y.W.).
\end{acknowledgments}

\appendix
\section{Relating Kerr angle with Hall conductivities
 in lattice systems}
\label{em}

In a rotationally-invariant system, the polar Kerr effect is associated with time-reversal symmetry breaking, and the Kerr angle $\theta_K$, measured in the experiments, is related to the antisymmetric component of the Hall conductivity $\sigma_{xy}-\sigma_{yx}$~(Refs.\ [\onlinecite{kerr_angle},\onlinecite{yakovenko}]).
In this Appendix we derive the expression for the Kerr angle $\theta_K$  for a system that is generally not rotationally invariant and prove that it is
 still proportional to antisymmetric off-diagonal component of the conductivity tensor $\sigma_{xy}-\sigma_{yx}$.

For a system that breaks both $C_4$ symmetry and time-reversal symmetry the conductivity tensor is given by
\begin{align}
\stackrel{\leftrightarrow}{\sigma}=\(\begin{array}{cc}\sigma_{xx}&\sigma_{xy}\\\ \sigma_{yx}&\sigma_{yy}\end{array}\),
\label{1}
\end{align}
 The the diagonal conductivities $\sigma_{xx}$ and $\sigma_{yy}$ are not necessarily the same and the Hall conductivities $\sigma_{xy}$ and $\sigma_{yx}$ generally have both a symmetric and antisymmetric components, i.e,
 $\sigma_{xy} + \sigma_{yx}$ and  $\sigma_{xy} - \sigma_{yx}$ are both non-zero.

\subsection{Kerr angle in a $C_4$ invariant system} 

For a lattice system which preserves $C_4$ invariance, $\sigma_{xx}=\sigma_{yy}$ and $\sigma_{xy}=-\sigma_{yx}$. The ``traditional" way~\cite{white} to define the Kerr angle is the following. One shines a linearly polarized light onto the surface. The incident linearly polarized light can be thought of as a superposition of left- and right-circularly polarized lights.
The two circularly polarized lights are eigenmodes for this rotational invariant system and
 under
 reflection they
  are rotated differently, with the angular difference
   $\Delta\phi$. When they are combined back together, the resulting new linearly-polarized light will be rotated by $\theta_{K}=\Delta\phi/2$. The relative rotation $\Delta \phi$ can expressed as
\begin{align}
\Delta\phi = \Imm\(\frac{\tilde r_-}{\tilde r_+}-1\),
\end{align}
where $\tilde r_{\pm}$ are complex reflectivities for left- and right-modes, respectively. From electrodynamics we have $\tilde r_\pm=(\tilde n_{\pm}-1)/(\tilde n_{\pm}+1)$, where $\tilde n_{\pm}$ are the complex indices of refraction.
Using this relation, we express the Kerr angle as
\begin{align}
\theta_K&=\Delta\phi/2=\frac{1}{2}\Imm\(\frac{\tilde r_-}{\tilde r_+}-1\)\nonumber\\
&=\Imm\( \frac{\tilde n_- -\tilde n_+}{\tilde n_+ \tilde n_- -1}\).
\label{ywwed11}
\end{align}
 We also know from electrodynamics that the complex indices of refraction satisfy $\tilde n_{\pm}^2=\tilde\epsilon_{\pm}$, where $\tilde\epsilon_{\pm}$ is the complex dielectric constant. The  complex dielectric constant can in turn be written as $\tilde\epsilon_{\pm}=\epsilon+i\frac{4\pi}{\omega}\sigma_{\pm}$, where $\epsilon$ is the conventional dielectric constant, $\omega$ is the light frequency, and $\sigma_{\pm}\equiv\sigma_{xx}\pm i\sigma_{xy}$ are the eigenvalues of $\stackrel{\leftrightarrow}{\sigma}$. We obtain,
\begin{align}
\tilde n _{-}-\tilde n_{+}\approx\frac{4\pi}{\omega n}\sigma_{xy},
\label{ywwed12}
\end{align}
where we have assumed that $\tilde n _{\pm}\approx n$  are close in value and predominantly real. Plugging Eq.\ (\ref{ywwed12}) into Eq.\ (\ref{ywwed11}) we obtain,
\begin{align}
\theta_K=\frac{2\lambda}{c}\Imm\[\frac{\sigma_{xy}}{n(n^2-1)}\],
\end{align}
where $\lambda$ is the wavelength of the incident light, $c$ is the speed of light, and $\sigma_{xy}=\sigma_{xy}(\omega)=-\sigma_{yx}(\omega)$ is the Hall conductivity at the light frequency $\omega$.

\subsection{Kerr angle in a system with broken $C_4$ symmetry}

For a system with broken $C_4$ symmetry the above analysis is inadequate. If we simply follow the same experimental method, we find that the Kerr angle defined this way will depend on both diagonal and off-diagonal element of the conductivity tensor, which means both rotational symmetry breaking and time-reversal symmetry breaking contributes to the ``Kerr angle". What is even worse, this ``Kerr angle" depends on the polarization direction of the incident light. This means this traditional experiment mixes contributions from rotation symmetry breaking and time-reversal symmetry breaking.

 In the experiment carried out by Kapitulnik group \cite{kapitulnik,xia}, the Kerr angle is measured by a more sophisticated technique. Instead of directly shining the linearly-polarized light unto the sample, they use a quarter-wave plate to transform the incident light into circularly polarized light. They   shine
  left and right polarized light on the surface and measure the relative phase shift between them off reflection. Since in two dimensions left and right circularly-polarized lights are connected by time-reversal, this experiment by definition measures time-reversal symmetry breaking and is immune to $C_4$ lattice symmetry breaking.

 We briefly recapitulate their setup. Linearly-polarized light with polarization $(1,1)$ passes
 through a phase modulator and is split equally into two perpendicular polarizations along the ``fast" axis $(1,0)$ and ``slow" axis $(0,1)$ axes of the phase modulator. We label these two modes as 1 and 2 and all the subsequent devices are carefully aligned such that modes 1 and 2 do not get mixed. The phase modulator introduces a spatial separation and relative phase shift $\phi_1$ between modes 1 and 2, which is oscillating in time with frequency $\omega_n$,
\begin{align}
\phi_1(t)=\phi_n\sin(\omega_n t).
\end{align}

The two modes exiting the phase modulator are incoherent, since the spatial separation between them is larger than the coherent length of the laser beam. These two modes are transmitted by a 10-m-long polarization-maintaining (PM) optical fiber. Both
 modes
 then go through a carefully aligned quarter-wave plate and become circularly-polarized incident
 modes
  on the sample surface. Off reflection, these two circular modes generally become distorted and go through the quarter-wave plate again. At this point,
the mode 1(2) that traveled along the {\it fast (slow)} axis now will have to travel along {\it slow (fast)} axis, in order to become coherent with each other again. Due to reflection,  mode 1 acquires a phase shift $\Delta\phi$ compared to mode 2, and we define it as twice the Kerr angle
\begin{align}
{\Delta\phi}\equiv2\theta_{K}.
\end{align}

To experimentally determine the value of this angle, they transmit these reflections back along the same path to the phase modulator, and then to the detector. They tune the time it takes to travel back to the phase modulator, such that after passing it again,
 modes 1 and 2 receive an additional relative phase shift of
  exactly the same amount.
Upon exiting the phase modulator, modes 1 and 2 have traveled exactly the same
 length and become coherent. They have an overall phase difference of $2\phi_n\sin(\omega_n t)+2\theta_K$. Labeling fast and slow axes as $(1,0)$ and $(0,1)$, we find that polarization of the resulting light exiting the phase modulator is $(-\exp[i(2\phi_n\sin(\omega_n t)+2\theta_K)],1)e^{i\omega t}$. Next this light passes a half-wave plate and a linear polarizer oriented at $45^\circ$, which projects the amplitude onto (1,1) direction, and then is picked up by the detector. The exiting amplitude is then
\begin{align}
E(t)&=E_n(t)e^{i\omega t}\nonumber\\
&=\frac{1}{\sqrt{2}}\{1-\exp[i(2\phi_n\sin(\omega_n t)+2\theta_K)]\}e^{i\omega t}.
\end{align}
 The detector will average out the fast oscillation $e^{i\omega t}$ with light frequency $\omega\sim 10^{14} {\rm Hz}$. The intensity sensed by the detector is varying
  at a much slower
   frequency $\omega_n$ and can be expressed as
\begin{align}
I(t)=& |E_n(t)|^2\nonumber\\
\propto&1-\cos2\theta_K\cos[2\phi_n\sin(\omega_nt)]\nonumber\\
&+\sin2\theta_K\sin[2\phi_n\sin(\omega_nt)].
\label{10}
\end{align}
Expanding this in first and second order harmonics in $\omega_n$ using the Jacobi-Anger identity,  we find that
\begin{align}
\theta_K=\frac{1}{2}\arctan\[\frac{J_2(2\phi_n)I_{\omega_n}}{J_1(2\phi_n)I_{2\omega_n}}\],
\end{align}
where $J$'s are Bessel functions, and $I_{\omega_n}$ and $I_{2\omega_n}$ are first and second Fourier components of $I(t)$.
This is Eq.\ (1) of Ref.\ [\onlinecite{xia}], and they used this equation to determine $\theta_K$.

Next we derive the expression of the Kerr angle $\theta_K\equiv\Delta\phi/2$, measured this way, in terms of conductivities. For the convenience of presentation, we introduce a new set of
 coordinates in which
 the fast axis is along $(1,1)$ and the slow axis is along $(-1,1)$.

Passing through the quarter-wave plate for the first time, the  linear modes $E_1=(1,1)$ and $E_2=(-1,1)$ become circular, $\tilde E_1=(1,i)$ and $\tilde E_2=(-1,i)$.
For a $C_4$ invariant system the conductivity tensor is antisymmetric. In this case the left- and right-circularly lights $\tilde E_1=(1,i)$ and $\tilde E_2=(-1,i)$ are eigenmodes of the system. Off reflection, both of them stay circularly polarized and merely receive a relative phase shift, which can be
  expressed in terms of reflectivities, like in Eq. (\ref{1}).

 Without rotation invariance, circular modes 1 and 2 are ${\it not}$ eigenmodes of $\stackrel{\leftrightarrow}{\sigma}$ and do {\it not} simply get multiplied by phase factors
   after
   reflection. For each of the modes, the way to proceed is to decompose it into eigenmodes of the conductivity tensor,
    study how they separately get rotated,
    and then
     combine the two phase-shifted eigenmodes back. The eigenmodes of the conductivity tensor (\ref{1}) are
\begin{align}
\tilde{E}_{\pm}=\(\begin{array}{c}1\\ \frac{\Delta\sigma\pm i \Sigma_{xy}}{\sigma_{xy}}\end{array}\),
\end{align}
where $\Delta\sigma\equiv(\sigma_{xx}-\sigma_{yy})/2$, and $i\Sigma_{xy}\equiv\sqrt{\sigma_{xy}\sigma_{yx}+\Delta\sigma^2}$. The corresponding eigenvalues of the conductivity tensor is found to be
\begin{align}
\sigma_{\pm}=({\sigma_{xx}+\sigma_{yy}})/{2}\pm i\Sigma_{xy}.
\end{align}

 Let us consider how mode 1 (the fast mode) behave after reflection. The wave function of mode 1 can be written as
\begin{align}
\tilde E_1=(1,i)=a\tilde {E}_{+}+b\tilde {E}_{-},
\end{align}
where
\begin{align}
a=\frac{\sigma_{xy}+\Sigma_{xy}+i\Delta\sigma}{2\Sigma_{xy}},~~~b=\frac{-\sigma_{xy}+\Sigma_{xy}-i\Delta\sigma}{2\Sigma_{xy}}.
\label{4}
\end{align}
 When the two modes $\tilde {E}_{\pm}$  are reflected from the surface, they acquire a relative factor,
\begin{align}
\tilde E_1'=a\tilde E_++be^{i\Delta\tilde\phi}\tilde E_-,
\label{44}
\end{align}
 We assume that $\Delta\tilde\phi$ is small but {\it not} necessarily real.
  Using standard procedures we find
\begin{align}
i\Delta\tilde\phi\approx&\frac{\tilde r_{-}}{\tilde r_{+}}-1=\(\frac{\tilde n_{-}-1}{\tilde n_{-}+1}~\frac{\tilde{n}_++1}{\tilde{n}_+-1}\)-1\nonumber\\
\approx&\frac{2(\tilde n_{-}-\tilde n_{+})}{n^2-1}
\approx\frac{\tilde \epsilon_{-}-\tilde\epsilon_{+}}{n(n^2-1)}\nonumber\\
=&\frac{8\pi}{\omega}\frac{\Sigma_{xy}}{n(n^2-1)},
\label{5}
\end{align}
where $\tilde r_{\pm}$ is reflectivity, $\tilde n_{\pm}=\sqrt{\tilde\epsilon_{\pm}}$ is the index of refraction, and $\tilde \epsilon_{\pm}=\epsilon+ i\frac{4\pi}{\omega}\sigma_{\pm}$ is the complex dielectric constant, where $\omega$ is the frequency of the light. We have assumed in the second line that $\tilde n_{+}\approx\tilde n _{-}\approx n$.
Subsequently, the reflected mode 1 goes through the same quarter-wave plate for the second time,
 and, as a result,  $\tilde E_{\pm}$
 gets converted into
\begin{align}
{E}_{\pm}=\(\begin{array}{c}1\\ \frac{i\Delta\sigma\mp  \Sigma_{xy}}{\sigma_{xy}}\end{array}\).
\label{6}
\end{align}
At this step, the polarization of mode 1 becomes
\begin{align}
E_1'=a E_{+}+ b e^{i\Delta\tilde\phi} E_{-}.
\label{66}
\end{align}
Combining results from Eqs.\ (\ref{4},\ref{6},\ref{66}), we find
\begin{align}
E_{1}'=&\frac{e^{i\Delta\frac{\tilde\phi}{2}}}{\Sigma_{xy}}\bigg [\Sigma_{xy}\cos\Delta\frac{\tilde\phi}{2}+i(\sigma_{xy}-i\Delta\sigma)\sin\Delta\frac{\tilde\phi}{2},\nonumber\\
&\bigg(\Delta\sigma+i\frac{\Sigma_{xy}^2+\Delta\sigma^2}{\sigma_{xy}}\bigg)\sin\Delta\frac{\tilde\phi}{2}-\Sigma_{xy}\cos\Delta\frac{\tilde\phi}{2}\bigg].
\end{align}
We recall that mode 1 traveled through the {\it fast} axis of the phase modulator and PM cable, and only its component along $\it slow$ axis now can travel back to coherently interfere with  the counterpart of mode 2.
 Projecting mode 1
 onto the slow axis $(-1,1)$, we find that
\begin{align}
E_{1}''=&-\frac{e^{i\Delta\frac{\tilde\phi}{2}}}{\Sigma_{xy}}\bigg[\Sigma_{xy}\cos\Delta\frac{\tilde\phi}{2}\nonumber\\
&-i\frac{\sigma_{xy}-\sigma_{yx}}{2}\sin\Delta\frac{\tilde\phi}{2}\bigg ]\times(-1,1),
\label{7}
\end{align}
where we have used $\Sigma_{xy}^2=-\sigma_{xy}\sigma_{yx}-\Delta\sigma^2$.
It is easy to verify that for a special case with $C_4$ symmetry, $E_1''=(1,-1)$ as expected.

The same consideration applies equally to mode 2, and its coherent part is
\begin{align}
E_{2}''=&-\frac{e^{i\Delta\frac{\tilde\phi}{2}}}{\Sigma_{xy}}\bigg[\Sigma_{xy}\cos\Delta\frac{\tilde\phi}{2}\nonumber\\
&+i\frac{\sigma_{xy}-\sigma_{yx}}{2}\sin\Delta\frac{\tilde\phi}{2}\bigg ]\times(1,1).
\label{8}
\end{align}

The relative phase shift between $E_2''$ and $E_1''$ is defined as twice the Kerr angle. Using (\ref{7}) and (\ref{8}) and assuming that $\Delta\tilde\phi$ and $\theta_K$ are small we
 obtain
\begin{align}
1+2i\theta_K\approx\exp{(2i\theta_K)}=\frac{|E_2''|}{|E_1''|}\approx1+\(\frac{\sigma_{xy}-\sigma_{yx}}{2\Sigma_{xy}}\)i\Delta\tilde\phi.
\end{align}
We are only interested in the real part of $\theta_K$. Using the result from Eq.\ (\ref{5}), we find that
\begin{align}
\theta_K=\frac{\lambda}{c}\Imm\[\frac{\sigma_{xy}-\sigma_{yx}}{n(n^2-1)}\],
\label{9a}
\end{align}
where the Hall conductivities are taken at the light frequency $\sigma_{xy}=\sigma_{xy}(\omega)$.
We see that $\Sigma_{xy}$, which contains the diagonal part of the conductivity tensor, gets canceled out.
It is
 easy
 to verify that
 $\sigma_{xy}(\omega)-\sigma_{yx}(\omega)$ is invariant under
  rotations in the xy plane.
   Therefore, Eq.\ (\ref{9}) is independent on the polarization of the incident light or the choice of coordinate system $x$ and $y$.

\section{Derivation of Eq.\ (\ref{ywsun1})}\label{app:1}
In this appendix we provide a more rigorous derivation of Eq. (\ref{ywsun1}). We start with Eq. (\ref{ywsun5}),
\begin{align}
\gamma(\omega_m)=&T\sum_{\Omega_m'}\frac{1}{|\Omega_m'|}\frac{\omega_m}{(\Omega_m')^2-\omega_m^2/4}.
\label{ywsun7}
\end{align}
We convert
 the frequency
 summation into a contour integral
\begin{align}
\gamma({\omega_m})=-\frac{1}{2\pi i}\oint_{C}\frac{dz}{e^{\beta z+\delta}+1}\frac{1}{|-iz|}\frac{\omega_m}{z^2+\omega_m^2/4},
\label{ywsun8}
\end{align}
where $\beta=1/T$.
We added
 $\delta\to0$ to the argument of the exponent to avoid unphysical divergencies (see below).

The integration contour $C$ is shown in Fig.\ \ref{cont}. We can equivalently replace $C$ with three contours: $C_1$ and $C_2$ are around two poles at $\pm\omega_m/2$, and $C_3$ is around a branch cut
along the real axis. We show these contours in Fig.\ \ref{cont}.

 \begin{figure}[t]
\includegraphics[width=\columnwidth]{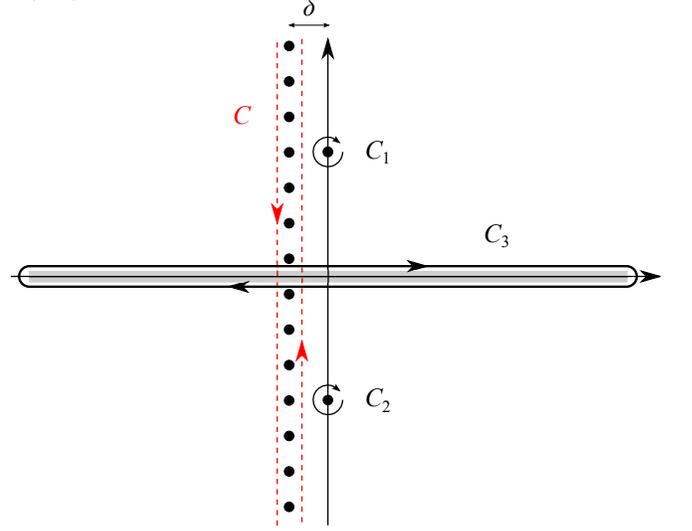}
\caption{The integration contour of Eq.\ (\ref{ywsun8}), in red dashed line, and the integration contour of Eqs.\ (\ref{ywsun111},\ref{ywsun112}), in black solid lines.}
\label{cont}
\end{figure}

Contour integrals
 over $C_1$ and $C_2$ yield, respectively,
\begin{align}
\gamma_1(\omega_m)=&\frac{-i}{e^{i\beta\omega_m/2+\delta}+1}\frac{2}{|\omega_m|}\nonumber\\
\gamma_2(\omega_m)=&\frac{i}{e^{-i\beta\omega_m/2+\delta}+1}\frac{2}{|\omega_m|}.
\label{ywsun111}
\end{align}
Using the fact that $\omega_m=2m\pi T$, we find that these two terms cancel out.
 The presence of $\delta$  is essential here as without it both $\gamma_1$ and $\gamma_2$ would diverge.
  To obtain $\gamma(\omega_m)$ we then only need to integrate over the branch cut
   along the contour $C_3$. We find
\begin{align}
\gamma(\omega_m)=&\frac{-1}{2\pi i}\int_{-\infty}^{\infty}\frac{dz}{e^{\beta z}+1}\frac{\omega_m}{z^2+\omega_m^2/4}\nonumber\\
&\times\[\frac{1}{\sqrt{-(z+i0)^2}}-\frac{1}{\sqrt{-(z-i0)^2}}\]\nonumber\\
=&\frac{1}{\pi}\int_{-\infty}^{\infty}\frac{dz}{e^{\beta z}+1}\frac{\omega_m}{z^2+\omega_m^2/4}\times\frac{1}{z}\nonumber\\
=&-\frac{1}{2\pi}\int_{-\infty}^{\infty}\frac{dz~\tanh(\beta z/2)}{z}\frac{\omega_m}{z^2+\omega_m^2/4},
\label{ywsun112}
\end{align}
where in the last step we have used $f(z)\equiv1/(e^{\beta z}+1)=[1-\tanh(\beta z/2)]/2$.
Evaluating this integral in the limit $\beta\omega_m\gg1$, we find,
\begin{align}
\gamma(\omega_m)=-\frac{4}{\pi\omega_m}\log\frac{|\omega_m|}{T}.
\end{align}
This is Eq.\ (\ref{ywsun1}) in the main text.

\end{document}